\documentstyle[12pt,epsfig,rotate]{article}

\newcommand{\degr}{\mbox{$^\circ$}}
\newcommand{\degC}{\mbox{$^\circ$C}}

\newcommand{\be}{\begin{equation}}
\newcommand{\ee}{\end{equation}}
\newcommand{\Ber}{\mbox{$^{7}$Be}}
\newcommand{\Bor}{\mbox{$^{8}$B}}

\newcommand{\Chl}{\mbox{$^{37}$Cl}}
\newcommand{\Arg}{\mbox{$^{37}$Ar}}
\newcommand{\Gal}{\mbox{$^{71}$Ga}}
\newcommand{\Ger}{\mbox{$^{71}$Ge}}
\newcommand{\nuster}{\mbox{$\nu_{s}$}}
\newcommand{\nue}{\mbox{$\nu_{e}$}}
\newcommand{\numu}{\mbox{$\nu_{\mu}$}}
\newcommand{\nutau}{\mbox{$\nu_{\tau}$}}

\newcommand{\nueb}{\mbox{$\bar{\nue}$}}

\newcommand{\goesto}{\mbox{$\rightarrow$}}
\def\percent{\%}
\def\dollar{\$}
\def\Cerenkov{$\check{\textrm{C}}$erenkov}

\begin{document}
\topmargin 0pt
\oddsidemargin=-0.4truecm
\evensidemargin=-0.4truecm
\pagenumbering{roman}

\begin{titlepage}
\vspace*{-2.0cm}
\begin{flushright}
%\emph{\today}\\
\end{flushright}
\vspace*{4.5cm}

\begin{center}
{\huge \textbf{Neutrino Physics}}

{\huge \textbf{\&}}

{\huge \textbf{The Solar Neutrino Problem}}

\vspace{1.0cm}
{\Large \emph{Andrew Lowe}}
\vspace{0.05cm}

%{\emph{Physics Department, Royal Holloway, University of London}}\\
\end{center}
\vglue 0.8truecm

%\begin{abstract}

%\end{abstract}

\end{titlepage}
\tableofcontents

\newpage
\pagenumbering{arabic}
\section{Introduction}
Neutrinos are weakly interacting neutral particles that travel essentially at the speed of light\footnote{This can be inferred from the measured arrival times of photons and neutrinos from Supernova~1987A.}~\cite{Stodolsky} and have intrinsic angular momentum of $\frac{1}{2}\hbar$~\cite{Pauli}. There are three distinct types, or flavours, of neutrino (\nue, \numu, and \nutau), corresponding to the three generations of leptons. There is strong evidence, namely, from studies of Z$^{0}$ production in e$^{+}$e$^{-}$ collisions at LEP~\cite{PDG}, that no additional \emph{light} neutrino types exist. Light neutrinos are those with mass less than half the Z$^{0}$ mass.

In 1930 Wolfgang Pauli hypothesised the existence of neutrinos to account for the apparent non-conservation of energy in beta-decay~\cite{Pauli}. Four years later Enrico Fermi included Pauli's hypothesised particle in a comprehensive theory of radioactive decay, and dubbed the new particle the \emph{neutrino}, meaning ``little neutral one'' in Italian~\cite{Fermi}. Neutrinos were subsequently discovered by Frederick Reines and Clyde L. Cowan, Jr., in an experiment performed in 1956 using a nuclear reactor~\cite{Reines}. Muon neutrinos were discovered later at Brookhaven National Laboratory~\cite{MuNuFound}, and were found to behave differently from those produced in association with electrons. Tau neutrinos were finally observed in July 2000 at Fermilab by the DONUT (Direct Observation of NU Tau) Collaboration~\cite{DONUT}.

In 1964 Ray Davis and John Bahcall published papers~\cite{Davis,Bahcall} proposing a 100,000 gallon (3.8$\times$10$^5$~litre) detector of tetrachloroethylene to measure the solar neutrino capture rate on chlorine, thereby testing directly the theory of nuclear energy generation in stars. The detector was built in the Homestake Gold Mine in Lead, South Dakota, and was the first experiment to detect solar neutrinos. The first results from the experiment were published in 1968~\cite{DavisRes,BahcallRes}. Currently seven experiments have observed solar neutrinos: Homestake, Kamiokande~II, Super-Kamiokande, SAGE, GALLEX, GNO, and SNO~\cite{BPtwoK,SNO}. Those that have published results (all those listed except SNO) have detected fewer neutrinos than predicted from solar models, the degree of deficiency being different in detectors of different types. The are, in fact, three solar neutrino problems, but the long-standing discrepancy between the standard solar model predictions of neutrino flux and observational results has become known as \mbox{\emph{the} Solar Neutrino Problem}.

\section{Homestake}
\subsection{Why Study Solar Neutrinos?}
Neutrinos are interesting to astronomers because they can reach us from otherwise inaccessible regions where photons are trapped. In the Sun the mean free path of a photon is about 10$^{-10}$ of the radius of the Sun; so short that it takes them several million years to diffuse from the solar interior to the surface. Neutrinos interact weakly with matter -- \mbox{cross-sections} for a 1~MeV electron neutrino or electron antineutrino are of order 10$^{-19}$~b~\cite{PPbook} -- and can escape directly from the solar core. The mean free path of neutrinos in matter are of order of several light-years. Neutrinos provide the only means of directly studying the processes that drive energy generation in the Sun's interior. They are the signatures of the fusion reactions that create them. About 3\percent\ of the Sun's energy is radiated in the form of neutrinos~\cite{JNB1}. The solar neutrino flux at the Earth's surface is about 10$^{11}$~cm$^{-2}$s$^{-1}$~\cite{JNB1}. Solar neutrinos shine down on us during the day, and shine up on us during the night!

Thermonuclear fusion in the Sun's interior is the only known process sufficient to provide energy for the time scales required by biological, geological, and astronomical evidence. Fossils of primitive organisms have been found that are older than one billion years old, the earliest rocks are 3.8 billion years old, and meteorites have been found that are 4.5 billion years old. Gravitational energy release could support solar luminosity for only ten million years. Chemical energy would last only ten thousand years~\cite{JNBbook}. The idea that the Sun's heat is produced by thermonuclear reactions that fuse light elements into heavier ones was suggested by Sir Arthur Eddington in 1920~\cite{Eddington} and expanded by Hans Bethe in his epochal 1939 paper~\cite{Bethe}.

The Sun is a typical main sequence star, calmly burning hydrogen without violent or rapid evolution. Energy generation in main sequence stars proceeds via the proton-proton ($pp$) chain or the carbon-nitrogen-oxygen (CNO) cycle, two distinct sequences of reactions. In his 1939 paper, Bethe favoured solar energy generation via the CNO cycle. The hydrogen burning process can be represented symbolically by the relation
\begin{eqnarray*}
4{^{1}H} \rightarrow {^{4}He} + 2{e^{+}} + 2\nu_{e} + 25~MeV.
\end{eqnarray*}

The idea of using chlorine as a detector of neutrinos was suggested by Bruno Pontecorvo in 1946~\cite{Pontecorvo} and discussed in detail later by Luis W. Alvarez~\cite{Alvarez}. This lead Ray Davis, Jr., and Don S. Harmer to show, using a reactor and 3,000 gallon (1.1$\times$10$^{4}$~litre) tetrachloroethylene detector at Savannah River, that \nue\ and \nueb\ are different~\cite{JNB4}.

The experimental discovery in 1958 that the cross-section for the production of \Ber\ by the fusion of $^{3}$He and $^{4}$He was more than a thousand times larger than was previously believed~\cite{Holmgren} lead to the suggestion, by Willy Fowler and Al Cameron, that \Bor\ might be produced in sufficient quantities (from \Ber\ and protons) in the Sun to produce an observable flux of neutrinos from \Bor\ beta-decay~\cite{Fowler}. Prompted by this suggestion John N. Bahcall demonstrated in 1963 that the capture rate of \Bor\ neutrinos on chlorine is enhanced by a factor of about 20 by transitions to excited states of argon~\cite{Bahcall}. Bahcall had also calculated the event rate to be expected in a 100,000 gallon tank of tetrachloroethylene.

On the basis of Bahcall's calculations, and Davis' experience at Savannah River, Davis suggested the chlorine experiment that eventually located at the Homestake Mine. In 1964 Bahcall and Davis published papers~\cite{Bahcall,Davis} proposing the feasibility of building a solar neutrino detector ``...to see directly into the interior of a star and thus verify directly the hypothesis of nuclear energy generation in stars.''\cite{Bahcall}. The hypothesis being tested was that proton fusion is the origin of solar energy.

By 1964 it was widely accepted that the Sun shines almost entirely via the $pp$ chain of reactions, although at the time there was still no \emph{direct} evidence that thermonuclear fusion was responsible for solar energy generation. A measurement of the capture rate of \Bor\ neutrinos on chlorine would verify the hypothesis of solar energy generation via thermonuclear fusion and determine which sequence of reactions is dominant. Furthermore, because the \Bor\ neutrino flux depends approximately upon the 24th power of the central temperature of the Sun, an upper limit for the central temperature could be set~\cite{Bahcall,JNBbook}.

\subsection{The Experiment}
The \Chl\ detector was built deep underground to avoid background radiation from cosmic rays. The neutrino target was 2.2$\times$10$^{30}$ atoms of \Chl\ in the form of 100,000 gallons (3.8$\times$10$^5$~litre) of liquid tetrachloroethylene (C$_{2}$Cl$_{4}$), a common cleaning fluid. The tank containing the target was a cylinder about 6~m in diameter by 14.5~m in length, and was located 1.5~km underground in the Homestake Gold Mine in Lead, South Dakota. The total cost to excavate the cavity in the mine, build the tank, and purchase the liquid was \dollar0.6~million (in 1965). The experiment began in 1968.
\begin{figure}[!htbp]
\begin{center}
\mbox{\epsfig{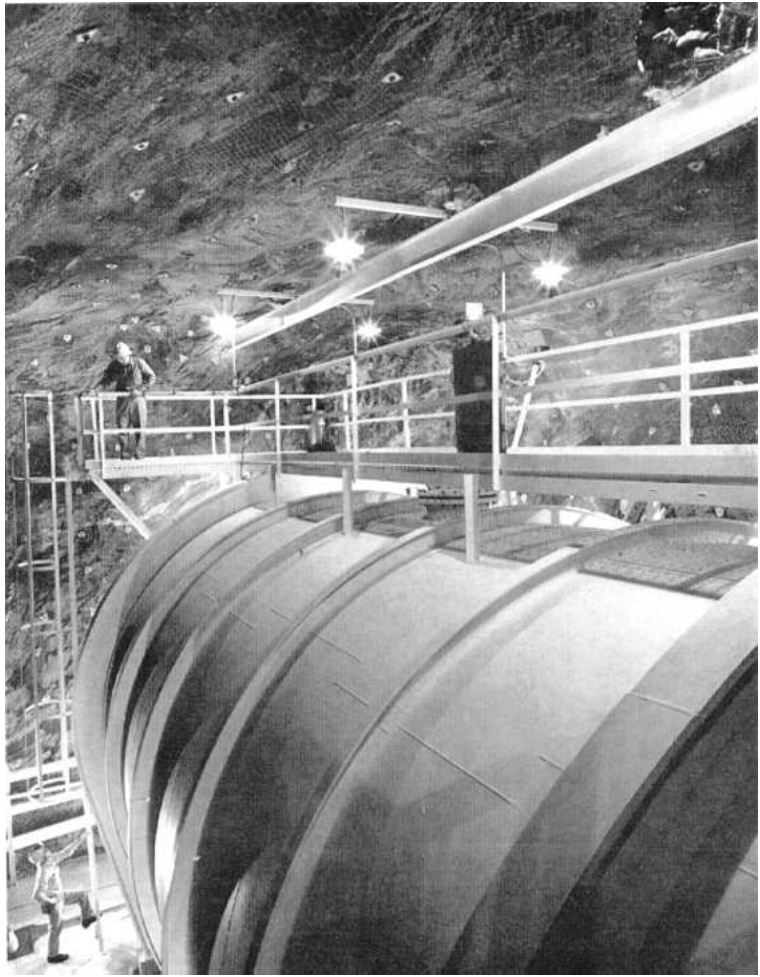}}
\caption{\small{The Homestake \Chl\ solar neutrino experiment. (From~\cite{JNB1}).}}
\label{Homestake}
\end{center}
\end{figure}
The neutrino absorption reaction that was used to detect solar neutrinos was
\begin{eqnarray*}
\nu_{e} + {^{37}Cl} \rightarrow e^{-} + {^{37}Ar}.
\end{eqnarray*}
The threshold energy for this reaction is 0.814~MeV. \Chl\ was chosen as a target because of its favourable physical and chemical characteristics:

\begin{itemize}
\item The absorption reaction has a relatively low threshold, permitting detection of all solar neutrino sources except the $pp$ neutrinos (see figure~\ref{snspec}).
\item The capture rate of \Bor\ neutrinos on \Chl\ is enhanced by more than an order of magnitude by transitions to excited states of \Arg.
\item 24.23\percent, a significant fraction, of naturally occurring chlorine atoms are \Chl.
\item Tetrachloroethylene is commonly used in dry-cleaning, and is manufactured on a large scale and relatively inexpensive. 
\item The capture process creates \Arg\ atoms with sufficient energy to break free from the parent molecules. These argon atoms dissolve in the liquid, and are easily removed by purging with helium gas.
\item The lifetime of~\Arg\ is 35 days. This is long enough that chemical extractions did not need to be performed too frequently, and short enough that many exposures could be obtained. Runs were typically of one to three months in duration.
\end{itemize}
The chemical process was relatively simple. This was important for performing the experiment and for convincing sceptics of the validity of the results. A small (about 0.1~cm$^{3}$) known amount of isotopically pure $^{36}$Ar (or  $^{38}$Ar), a non-radioactive isotope of argon, was added to the tank and allowed to dissolve in the liquid before each exposure. The amount of argon recovered at the end of each run gave a direct measure of the efficiency of the extraction of \Arg\ produced by neutrino capture.

The argon was removed from the liquid by bubbling large quantities (about 4$\times$10$^{5}$~litre) of helium gas through the system. The argon was separated from the helium gas by passing the gas through a condenser at 241~K, a molecular sieve at room temperature, and finally a charcoal trap maintained at 77~K. Purging the tank for 22 hours would usually recover 95\percent\ of the argon, which was retained on the charcoal trap. The charcoal trap was heated in order to release the argon that had condensed on it. The argon was then purified and the volume of the recovered sample was measured, which allowed the recovery efficiency to be calculated. Isotopic composition of the sample was determined by mass spectrometry at a later time.

Davis and his collaborators have performed two additional tests demonstrating the high efficiency of the extraction of \Arg. In one, a small neutron source was placed in the centre of the tank and the amount of \Arg\ produced by the neutron source in the tank was measured. In the second test a measured number of \Arg\ atoms (500) were added to the tank, and the quantity recovered was measured. A test was also performed to demonstrate that the \Arg\ atoms produced by neutrino capture do not form molecules but instead become free neutral atoms.

The radioactive \Arg\ was counted by loading the sample of recovered argon in a small gas proportional counter. The sample was counted for about eight months to determine the background characteristics of the counter. The counters were calibrated every two months with a $^{55}$Fe source, and were upgraded continually during operation of the experiment. They were located underground in a laboratory at Homestake, and shielded from outside sources of background radiation by a lead shield~\cite{JNBbook}. The background was also minimised by using low radioactivity materials in the counting apparatus. The background rate in the counters, for runs made since 1984, was calculated to be 3.6 counts per year~\cite{JNBbook}.

The main source of background in the experiment was caused by deeply penetrating muons produced by cosmic rays in the atmosphere. These cause cascades of energetic pions, protons, and neutrons that produce \Arg~\cite{JNBbook}. Backgrounds from high-energy muons have been estimated by exposing 600 gallon (2271~litre) tanks of tetrachloroethylene at higher levels in the mine and then extrapolating production rates of \Arg\ to the lower level where the Homestake experiment was located. The background capture rate from high-energy muons was estimated to be 0.08$\pm$0.03~day$^{-1}$~\cite{JNBbook}. % <-- SNU's not defined yet!

\subsubsection{The Capture Reaction}
\begin{figure}[!htbp]
\begin{center}
\mbox{\epsfig{file=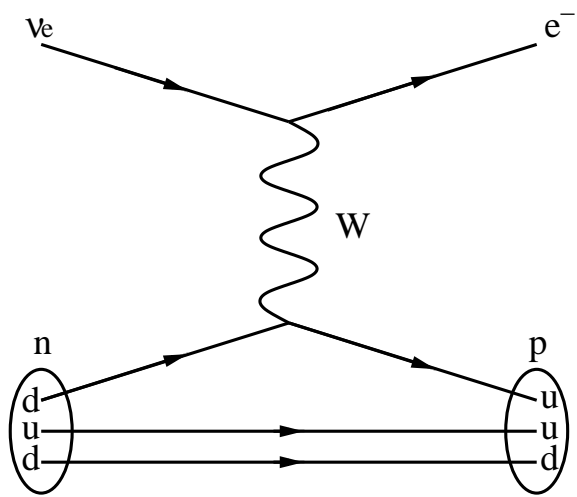,width=7cm,angle=270}}
\caption{\small{The neutrino capture process.}}
\label{capture}
\end{center}
\end{figure}

The neutrino capture reaction is
\begin{eqnarray*}
\nu_{e} + {^{37}Cl} \rightarrow e^{-} + {^{37}Ar}.
\end{eqnarray*}
This may also be written
\begin{eqnarray*}
\nu_{e} + n \rightarrow e^{-} + p.
\end{eqnarray*}
The process can be illustrated by means of a Feynman diagram (figure~\ref{capture}).

\noindent Hence it is clear that the charged current weak interaction is responsible for the underlying process in the neutrino capture reaction.

\subsection{Theoretical Expectations}
\subsubsection{The Standard Solar Model}
At this point it is appropriate to discuss in greater detail the solar model that the Homestake \Chl\ experiment was designed to verify. At the time that Bahcall and Davis published papers~\cite{Bahcall,Davis} proposing the feasibility of a chlorine experiment, most theorists favoured solar models in which the $pp$ chain was the primary source of energy generation, and that the CNO cycle is only significant for stars more massive and hotter than the Sun. However, there was no direct evidence that these processes were involved in solar energy generation, and certainly no proof that the $pp$ chain was the dominant process in the Sun.

\begin{figure}[!htbp]
\begin{center}
\mbox{\epsfig{file=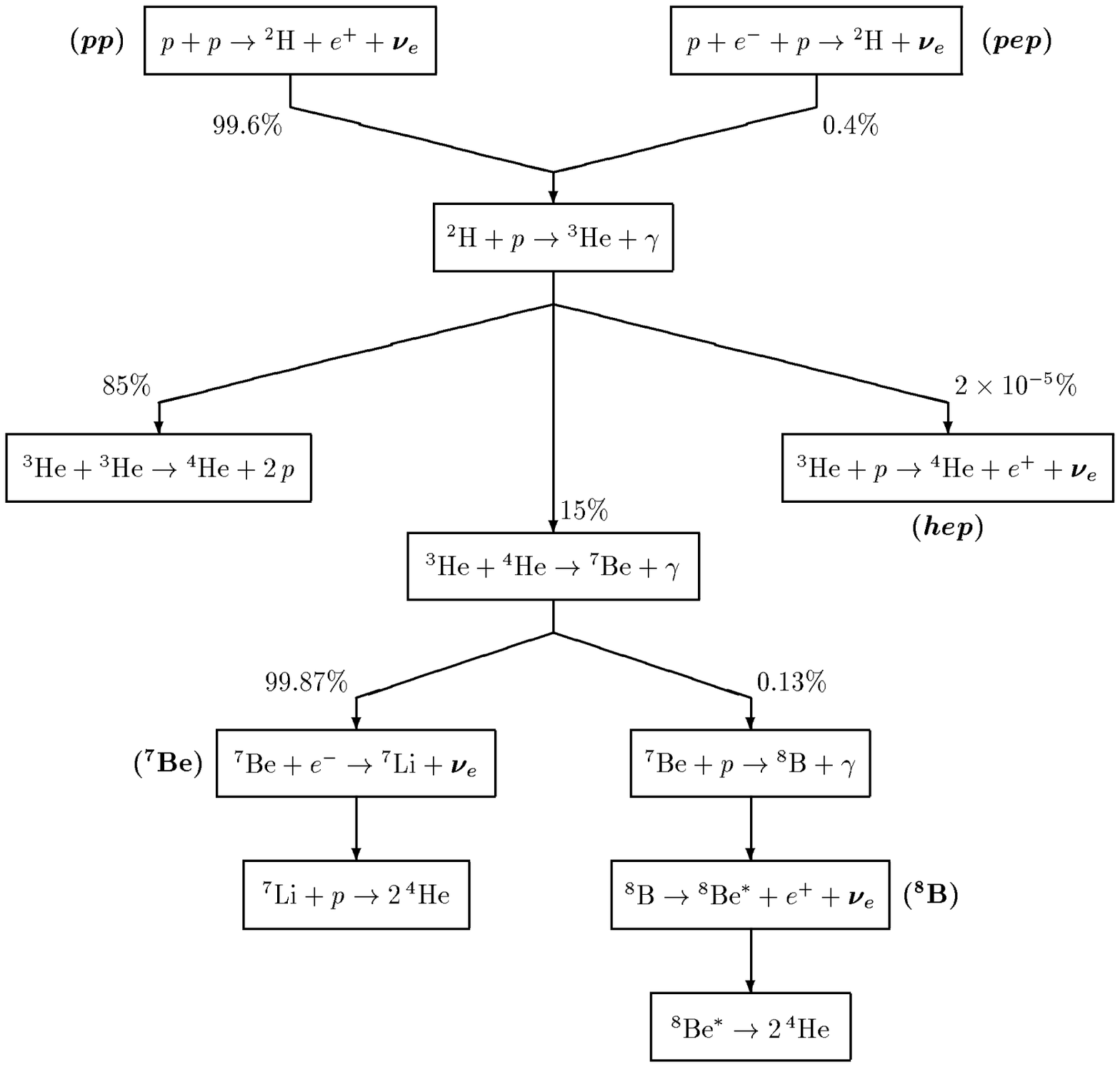,width=10cm}}
\caption{\small{The $pp$ chain. (From~\cite{Bilenky}).}}
\label{ppfig}
\vspace{1cm}
\mbox{\epsfig{file=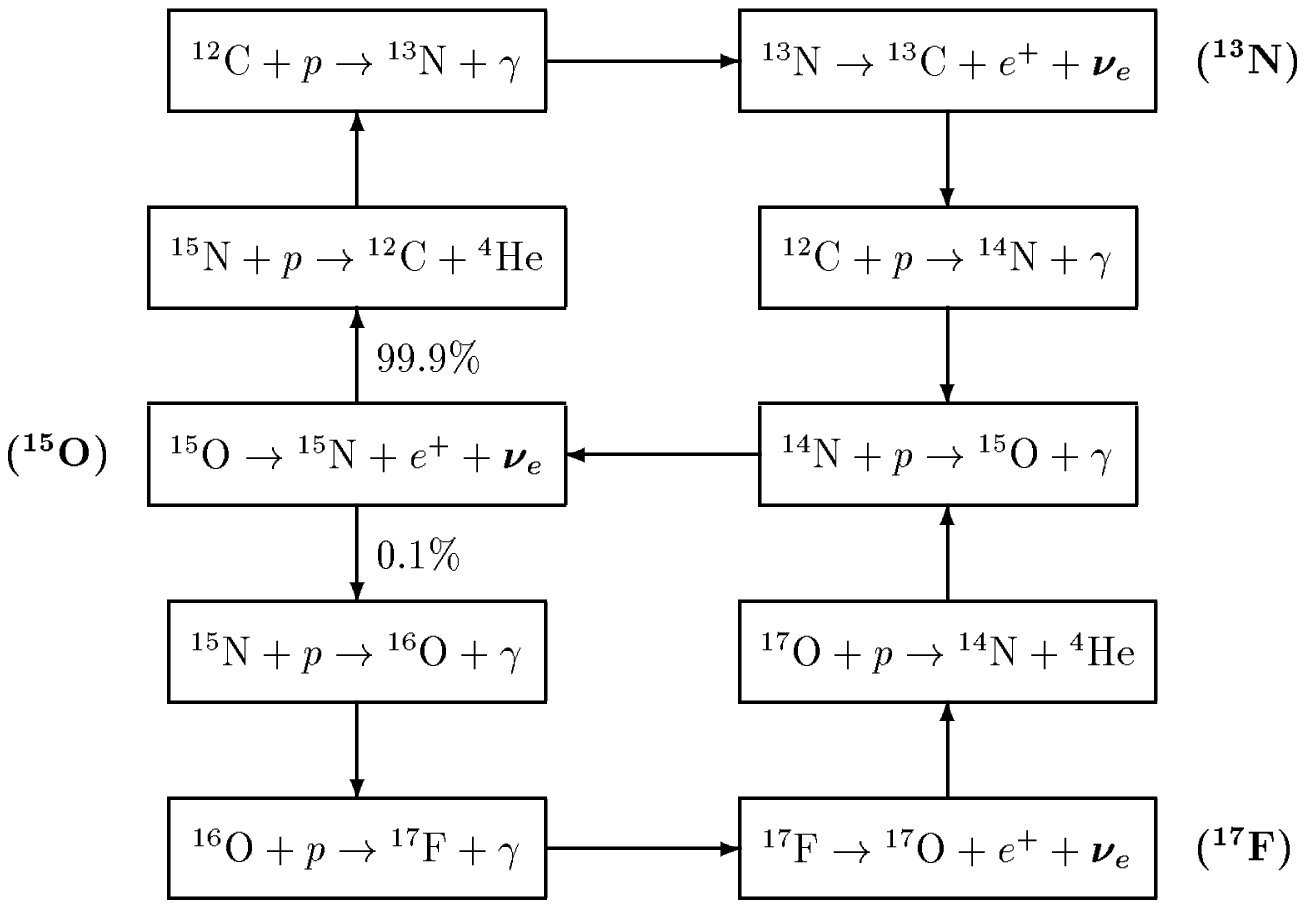,width=10cm}}
\caption{\small{The CNO cycle. (From~\cite{Bilenky}).}}
\label{CNOfig}
\end{center}
\end{figure}

The reactions that occur in the $pp$ chain and CNO cycle are shown in figure~\ref{ppfig} and figure~\ref{CNOfig}. Each completed conversion of four protons into an alpha particle, two positrons, and two electron neutrinos is known as a \emph{termination} of the $pp$ chain or CNO cycle. The percentages shown are the fraction of terminations of the $pp$ chain or CNO cycle in which each reaction occurs. Neutrinos are produced in six of the reactions of the $pp$ chain, and three of the reactions of the CNO cycle. These reactions are labelled on the diagram by the text in parentheses. The reactions take their names from the reactants. For example, the reactants in the ``pep'' reaction are proton, electron, and proton. Note the relative rarity of the \Bor\ neutrinos. However, it is these neutrinos which dominated the predicted capture rate in the Homestake \Chl\ experiment, because only these neutrinos are energetic enough to excite transitions between the ground state of \Chl\ and excited states of \Arg.

\begin{figure}[!htbp]
\begin{center}
\mbox{\epsfig{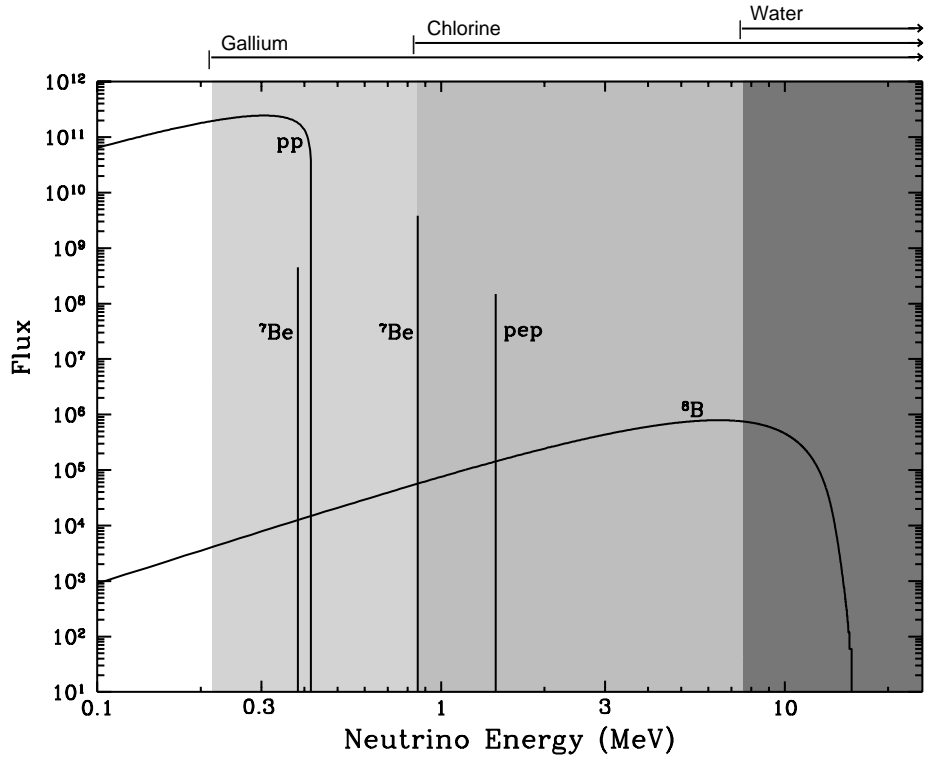}}
\caption{\small{Solar neutrino energy spectrum. Neutrino fluxes from continuum sources (like $pp$ and \Bor) are given in units of number per cm$^{2}$ per MeV at one astronomical unit, line fluxes (pep and \Ber) are given in number per cm$^{2}$ per second. Energy thresholds for the three different types of experiments are indicated above the figure. Neutrinos produced in the CNO cycle are not important energetically and have been omitted for clarity. (From~\cite{JNB2}).}}
\label{snspec}
\end{center}
\end{figure}

Figure~\ref{snspec} shows the energy spectrum of solar neutrinos predicted by the standard solar model. Notice that line spectra are predicted for the \Ber\ and $pep$ neutrinos. In the reactions that produce these neutrinos (see figure~\ref{ppfig}) there are only two final particles ($^{2}$H and neutrinos, or $^{7}$Li and neutrinos). The laws of conservation of energy and momentum require that for two final state particles, the ratio of the kinetic energies of the particles is equal to the inverse ratio of their masses. The energy must always divide this way, therefore the \Ber\ and $pep$ neutrinos are mono-energetic. With three or more final particles, such as those produced in the other reactions in the $pp$ chain, the total energy of the reaction can be shared among the final state particles many ways, resulting in continuous energy spectra for the neutrinos produced in these reactions.

\subsubsection{Predicted Capture Rates}
For convenience, neutrino capture rates are expressed in terms of a special unit; the solar neutrino unit, or \emph{SNU} (pronounced ``snew''). A SNU is the product of neutrino flux times neutrino cross section. One SNU equals 10$^{-36}$ captures per target atom per second. For the Homestake \Chl\ experiment, one atom of \Arg\ produced per day in the tank equals 5.35~SNU~\cite{JNBbook}.

Two pieces of information are needed to calculate neutrino capture rates: the solar neutrino fluxes at the Earth, and the cross sections for the neutrino capture reaction. One might expect that the solar interior is too complex to predict neutrino fluxes, but that is not quite the case. At the temperatures and densities at the Sun's core, the matter is fully ionised and is a close approximation of a perfect gas, hence the equation of state in the solar interior is relatively simple and can be calculated to sufficient accuracy to predict neutrino fluxes~\cite{JNBbook}. Neutrino capture cross sections can be determined empirically in nuclear physics experiments, and the nuclear properties of \Chl\ and \Arg\ are now well known. In 1964 scant information was available to Bahcall, but four years later improvements in the measured nuclear reaction rates allowed Bahcall to predict more accurately the capture rates to be expected for a \Chl\ detector. The predicted capture rates, calculated by Bahcall and collaborators in 1968~\cite{BahcallRes}, are as follows:
\begin{itemize}
\item If the CNO cycle is the dominant source of the Sun's energy, the capture rate was estimated to be 35~SNU~\cite{JNB1,JNB5}. A recently calculated value is 28~SNU~\cite{JNBbook}.
\item If the $pp$ chain is the dominant source of the Sun's energy, the capture rate, based on ``standard'' parameters, was estimated to be 6~SNU~\cite{JNB1}. The current value of the capture rate is calculated to be 7.6$^{+1.3}_{-1.1}$~SNU (1$\sigma$ errors)~\cite{BPtwoK}.
\end{itemize}

\subsection{The Results}
The first results from the Homestake \Chl\ detector were published by Davis and his colleagues in 1968~\cite{DavisRes}, and demonstrated that the capture rate of solar neutrinos on \Chl\ was less than 3~SNU~\cite{JNB1}. Despite the discrepancy between the predicted capture rate and the observed value, the Homestake \Chl\ experiment was very successful. Firstly, neutrinos were observed with rates within a factor of a few of the predicted values. Secondly, the large capture rate expected if the CNO cycle was dominant was \emph{not} observed. The first results from Homestake clearly implied that less than 10\percent\ of the Sun's energy is generated by the CNO cycle~\cite{JNB1,BahcallRes}. It is now thought that the CNO cycle is responsible for less than 2\percent\ of the Sun's luminosity~\cite{JNBbook}.

The energy production rates for the $pp$ chain and CNO cycle are temperature dependant. In stars much more massive than the Sun energy generation is dominated by the CNO cycle because their interior temperatures are higher. Also, the flux of \Bor\ neutrinos that dominated the capture rate in the Homestake \Chl\ experiment depends approximately on the 24th power of the central temperature of the Sun. Calculations based on the observed neutrino capture rate determined the central temperature to be close to the predicted value of 16$\times$10$^{6}$~K~\cite{JNBEAA,JNBbook}.

The Homestake \Chl\ experiment had found direct evidence that thermonuclear fusion is responsible for energy generation in the Sun, thus achieving the goal proposed in 1964, and that the dominant sequence of reactions is the $pp$ chain.

The Homestake experiment is still taking data, and during the three decades of its operation it has continued to observe fewer neutrinos than predicted. The most recently published value for the observed neutrino capture rate was 2.56$\pm$0.23~SNU (1$\sigma$ error)~\cite{BPtwoK}; about one third of the neutrinos appear to be missing. The persistent discrepancy between the predicted and the observed results is commonly known as the solar neutrino problem. This is, in fact, the first of \emph{three} problems. We will meet the others in due course. % <-- First use of ``we''.

In 1978, after a decade of disagreement between theoretical predictions and experimental results, it was clear that the subject had reached an impasse and a new experiment was required. The Homestake \Chl\ experiment was primarily only sensitive to the \Bor\ neutrinos that are produced in only 2 of every 10$^{4}$ terminations of the $pp$ chain. An experiment was needed that could exclude certain explanations for the solar neutrino problem. To do this it would have to be sensitive to the low energy $pp$ neutrinos undetectable by a \Chl\ experiment. The only possibility appeared to be another radiochemical experiment, this time with \Gal. The energy threshold for a gallium detector would be low enough to enable the detection of $pp$ neutrinos (see figure~\ref{snspec}), but unfortunately gallium was expensive, and about three times the world's annual production of gallium would be required. Although Davis, Bahcall, and a number of interested colleagues tried to generate interest in a gallium experiment, it was never funded in the US, mostly because of disagreements over who had financial responsibility for the experiment. It was not until the 1990s that gallium experiments were performed.
%%%%%%%%%%%%%%%%%%%%%%%%%%%%%%%%%%%%%%%%%%%%%%%%%%%%%%%%%%%%%%%%%%%%%%%%%%%%%%%
\section{Kamiokande~II}
Grand Unified Theories (GUTs) predict a long but finite lifetime for the proton on the order of 10$^{33}$ years. This is much longer than age of the universe. However, if a very large number of protons are monitored, and if protons have a finite lifetime, it might be possible to observe proton decay. If protons were observed to decay this would be evidence of physics beyond the Standard Model.

In the early 1980s two experiments were built to search for proton decay: IMB and Kamiokande~I. IMB (Irvine, Michigan, Brookhaven) was located in the Morton-Thiokol salt mine near Fairport, Ohio, at a depth of 600~m. Kamiokande (Kamioka Nucleon Decay Experiment) was located in the Mozumi zinc mine in Kamioka-cho, Gifu, 1000~m underground in the Japanese Alps. Both experiments used water \Cerenkov\ detectors.

\Cerenkov\ radiation is emitted whenever a charged particle travels through a transparent medium faster than the speed of light in that medium. The radiation occurs mostly in the blue and ultraviolet region of the spectrum, and forms a cone of light of half angle $\Theta_{c}$, called the \Cerenkov\ angle, given by:
\begin{eqnarray*}
\cos \Theta_{c}=\frac{c}{v\,n}
\end{eqnarray*}
where $c$ is the speed of light in vacuum, $v$ is the speed of the particle, and $n$ is the refractive index of the medium. For water with $n$=1.33, the \Cerenkov\ angle has a maximum value of 42\degr. The phenomenon is analogous to the acoustic shock wave produced when an object travels faster than the speed of sound in air. In a water \Cerenkov\ detector, when the cone of light reaches the wall of the detector it forms a ring-shaped pattern. Examples can be seen in figure~\ref{rings}.

\begin{figure}[!htbp]
\begin{center}
\vspace{-1.5cm}
\mbox{\epsfig{file=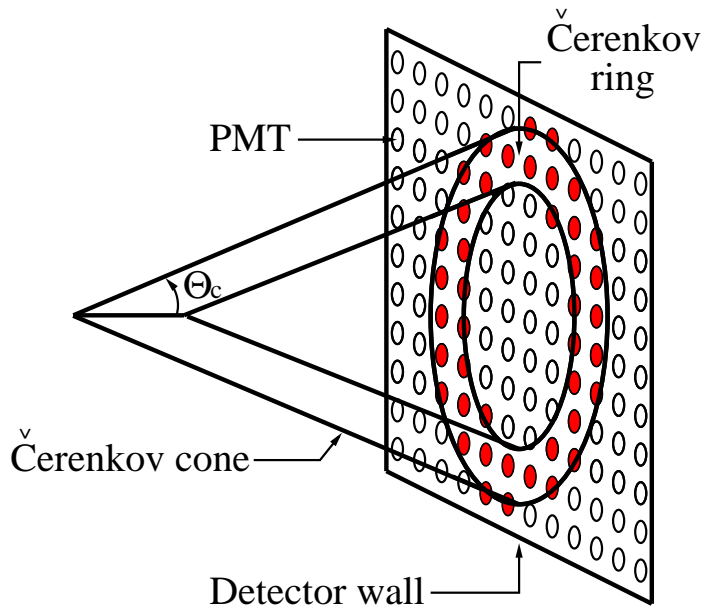,width=10cm,angle=270}}
\vspace{-0.5cm}
\caption{\small{The geometry of the \Cerenkov\ cone.}}
\label{cone}
\end{center}
\end{figure}

There are many good reasons for using water in a \Cerenkov\ detector. Water is cheap, abundant, can be purified to have low concentrations of radioactive contaminants, and can serve as both the detector and the target in neutrino-electron scattering experiments. A disadvantage is that one has to introduce energy cuts for the recoil electrons to suppress the background. For solar neutrino experiments, this means water \Cerenkov\ detectors are sensitive only to the $hep$ and \Bor\ neutrinos. For example, the threshold energy above which recoil electrons were counted in the Kamiokande~II detector was 7.5~MeV~\cite{JNBbook}.

\noindent IMB and Kamiokande looked for decays such as
\begin{eqnarray*}
p \rightarrow e^{+}\hspace{-10pt}&\pi^{0}&\\
&\pi^{0}& \rightarrow\;\gamma\gamma.
\end{eqnarray*}
In many models this particular decay mode is predicted to be dominant. It has a characteristic event signature, in which the electromagnetic shower caused by the positron is balanced against two showers caused by the gamma rays from the decay of the pion. The main background comes from neutrino interactions. Both experiments ran for about a decade, but no decays were observed, which set a lower limit for the proton lifetime~\cite{pdecay1,pdecay2}. Since neutrino interactions mimic certain expected proton decay channels, these detectors also began to study atmospheric neutrinos. 
% IMB technical specifications:
The IMB detector was a rectangular tank measuring 18~m by 17~m by 23~m filled with purified water, and instrumented on six sides with 2048 photomultiplier tubes (PMTs). The dominant reaction by which neutrinos were detected was antineutrino absorption by protons:
\begin{eqnarray*}
\bar{\nu_{l}} + p \rightarrow n + l^{+}.
\end{eqnarray*}
The cross section for this reaction is about two orders of magnitude larger than the cross section for neutrino-electron scattering. IMB was less sensitive to low-energy events than Kamiokande~II, and was not used to study solar neutrinos.
% Kamiokande~II technical specifications:
In 1985 the Kamiokande~I detector was upgraded to make it sensitive to low-energy events from solar neutrinos. Kamiokande~II was put into full-time service as a neutrino observatory in late 1986. The process by which neutrinos were detected at Kamiokande~II was neutrino-electron scattering. The water detector was contained in a cylindrical tank, 15.6~m in diameter by 16~m in height. The tank contained 3000~tons of pure water, but only the inner 380 tons was used for solar neutrino experiments because of stringent background requirements. The inner volume of water was a cylinder of radius 4.0~m and height 5.2~m, and was further inside the tank and therefore shielded from outside sources of radioactivity, such as gamma rays from the surrounding rock. Approximately 20\percent\ of the inner surface of Kamiokande~II was instrumented with 948 specially constructed PMTs.

\subsection{Neutrino Detection}
In water \Cerenkov\ detectors three different types of neutrino interactions occur: elastic scattering off electrons, quasi-elastic scattering off nucleons, and inelastic interactions with nucleons. These can further be classified as charged-current (CC) or neutral-current (NC) weak interactions.

\subsubsection{Neutrino-Electron Scattering}
\begin{figure}[!htbp] % Feynman diagrams.
\begin{center}
\vspace{-1cm}
\epsfig{file=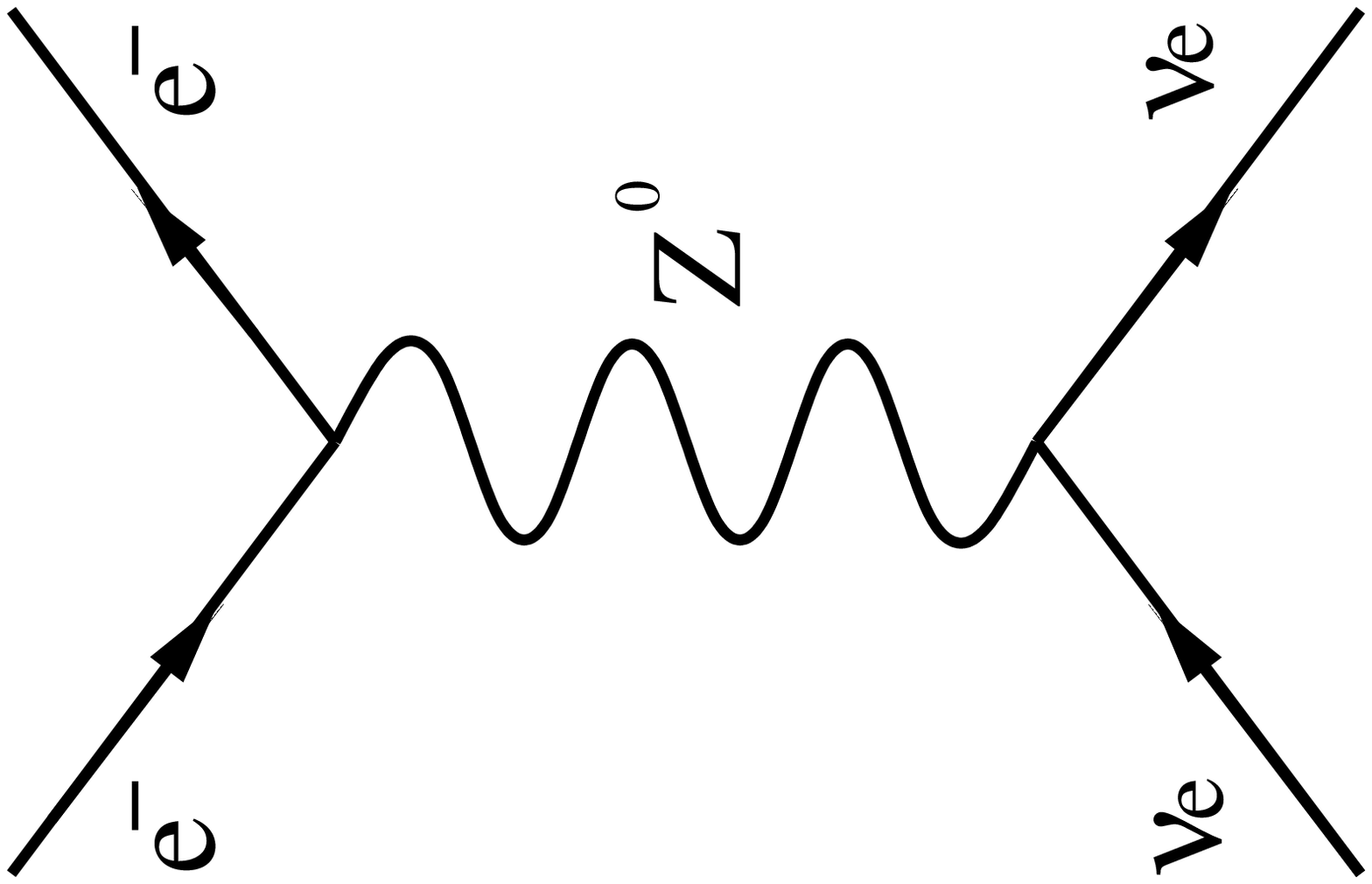,height=7cm,angle=270}
\epsfig{file=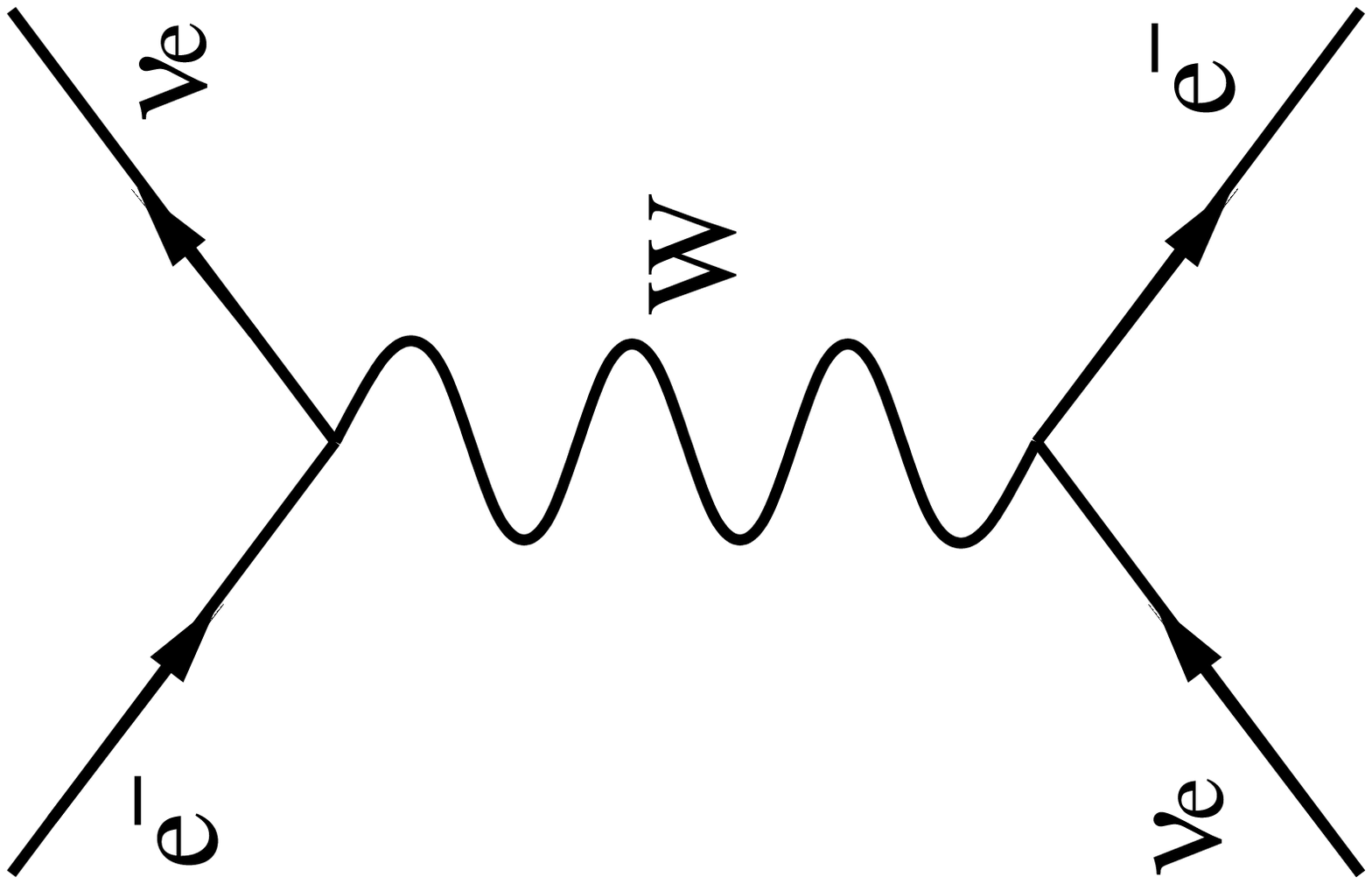,height=7cm,angle=270}
\vspace{-0.75cm}
\caption{$\nu_{e} e^{-} \rightarrow \nu_{e} e^{-}$}
\label{scat1}
%\end{figure}
%\begin{figure}[h] 
\vspace{-1cm}
\epsfig{file=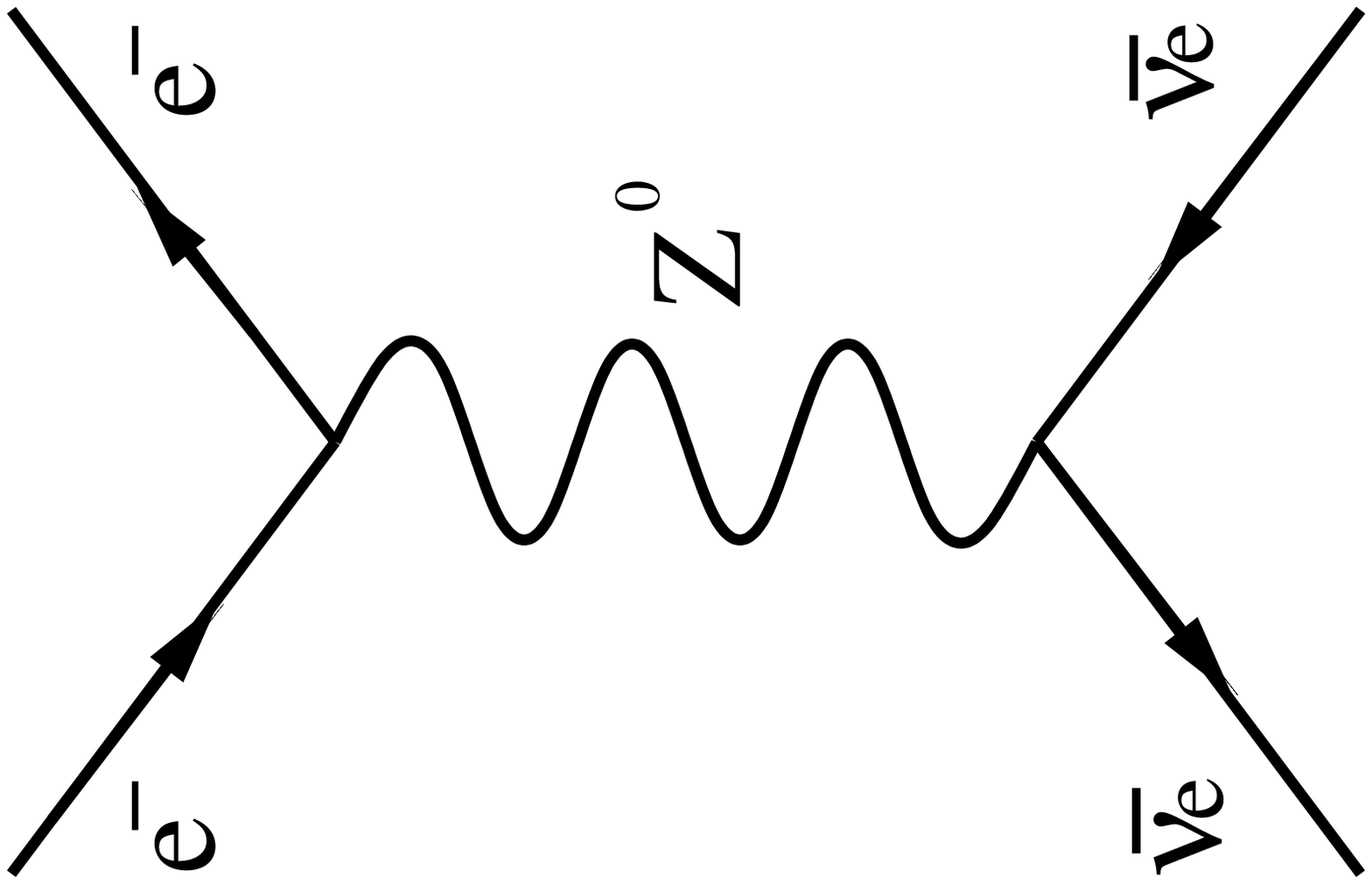,height=7cm,angle=270}
\epsfig{file=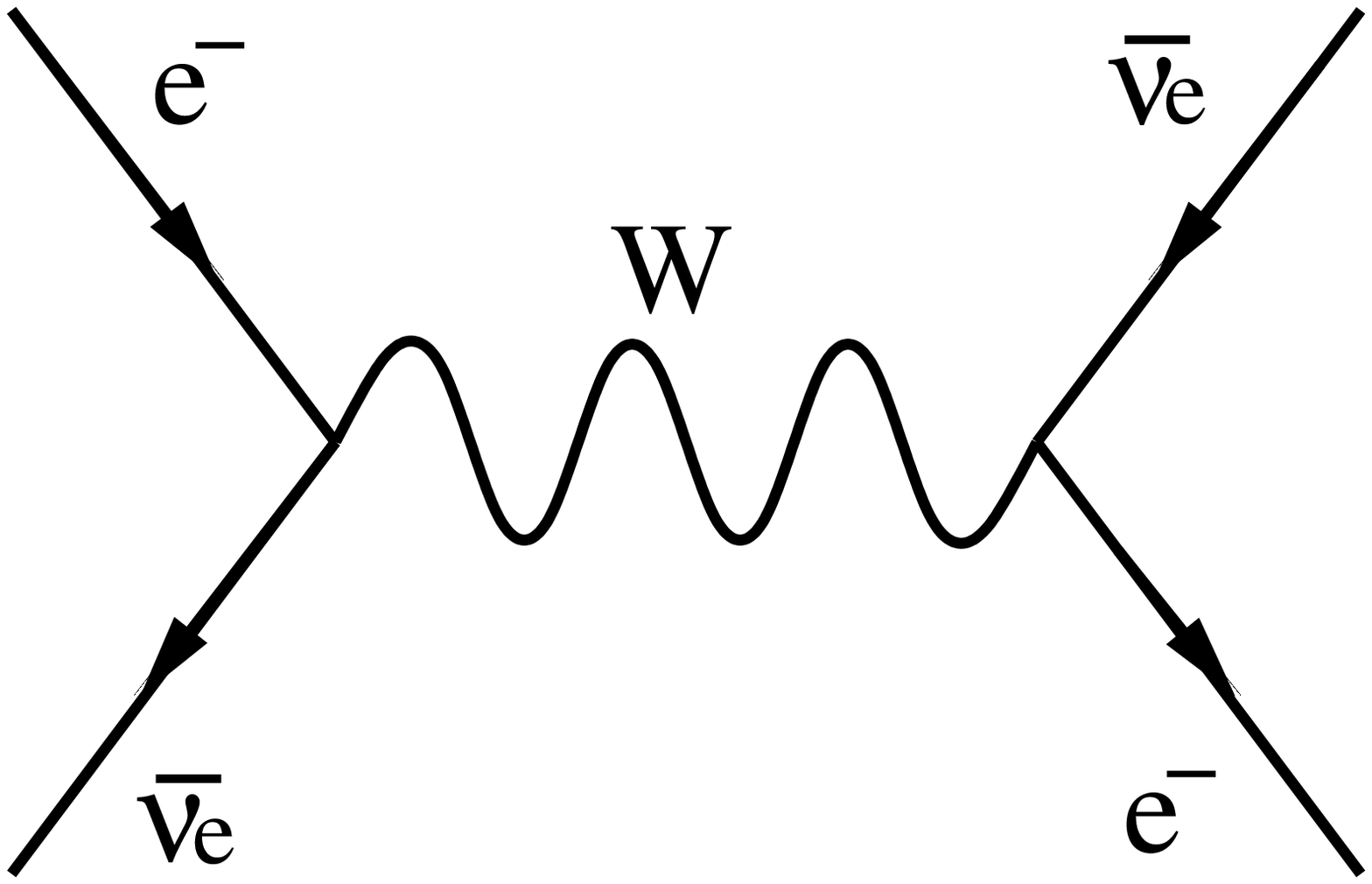,height=7cm,angle=270}
\vspace{-0.75cm}
\caption{$\bar{\nu_{e}} e^{-} \rightarrow \bar{\nu_{e}}  e^{-}$}
\label{scat2}
\end{center}
%\end{figure}
%\begin{figure}[h]
\begin{center}
\vspace{-1cm}
\epsfig{file=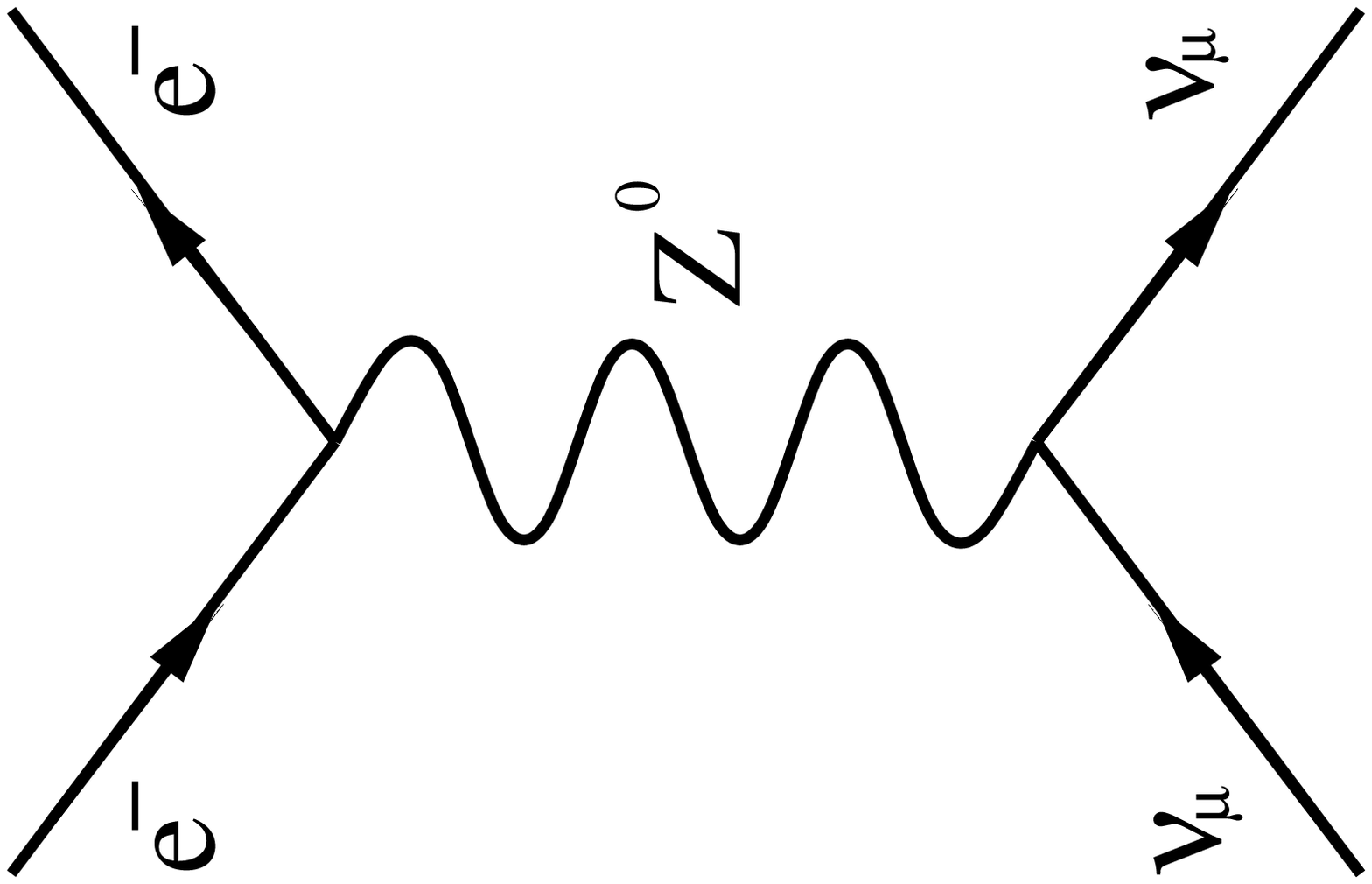,height=7cm,angle=270}
\vspace{-0.75cm}
\caption{$\nu_{\mu} e^{-} \rightarrow \nu_{\mu} e^{-}$}
\label{scat3}
\end{center}
%\end{figure}
%\begin{figure}[h]
\begin{center}
\vspace{-1cm}
\epsfig{file=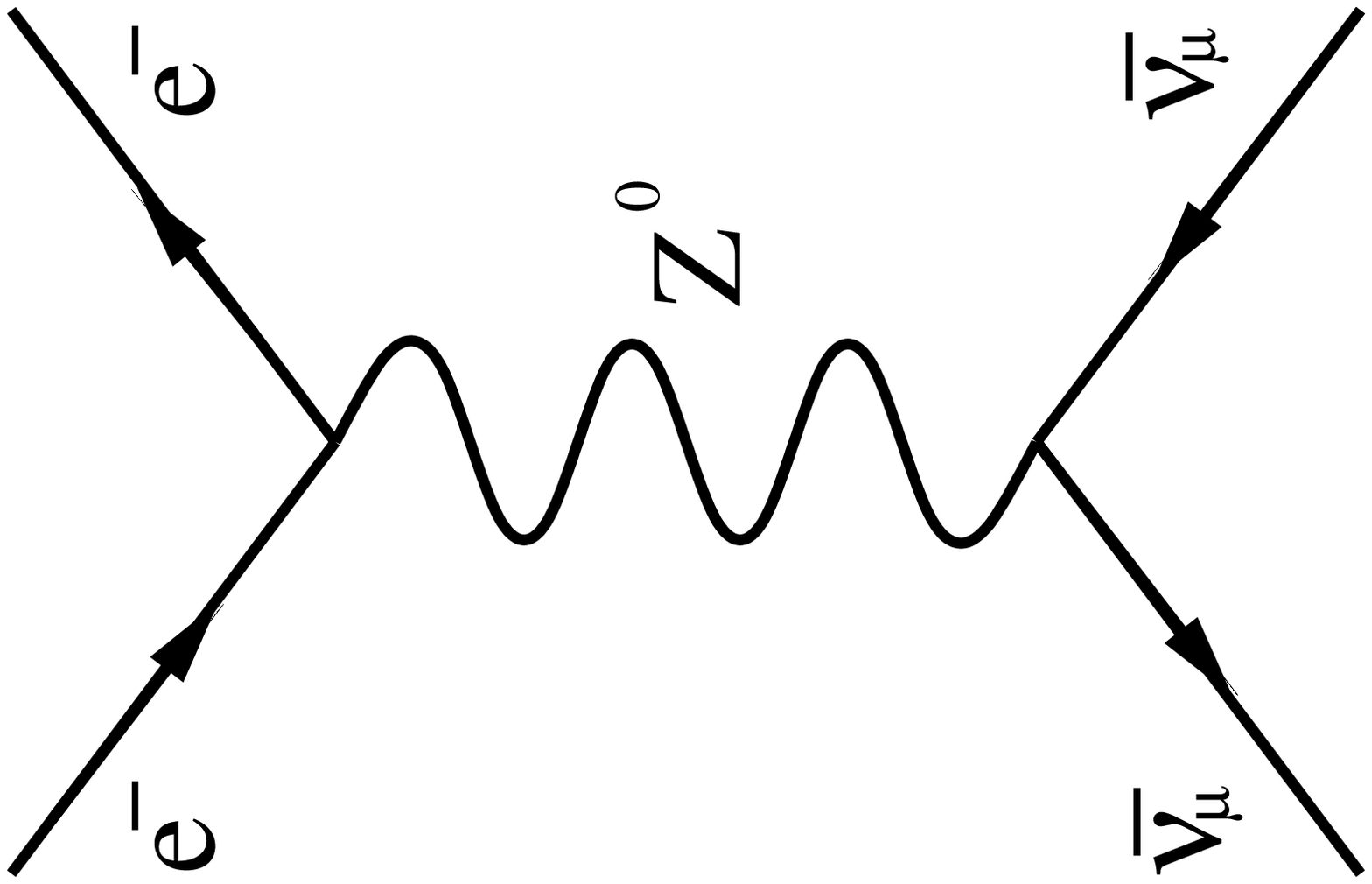,height=7cm,angle=270}
\vspace{-0.75cm}
\caption{$\bar{\nu_{\mu}} e^{-} \rightarrow \bar{\nu_{\mu}} e^{-}$}
\label{scat4}
\end{center}
\end{figure}

The elastic neutrino-electron scattering process can be represented symbolically by the relation
\begin{eqnarray*}
\nu_{l} + e^{-} \rightarrow \nu_{l} + e^{-}.
\end{eqnarray*}
Some tree-level Feynman diagrams for this process are shown in figure~\ref{scat1}, \ref{scat2}, \ref{scat3}, and \ref{scat4}. The advantages of a water \Cerenkov\ detector using the neutrino-electron scattering interaction over a radiochemical detector are threefold. Firstly, the exact arrival time of the incident neutrinos can be determined. In a radiochemical experiment one would only be able to determine the number of neutrinos detected since the last chemical extraction. Secondly, the direction of the incident neutrinos can be inferred from the \Cerenkov\ ring geometry. This is because recoil electrons are scattered in almost the same direction as the neutrinos. Reconstruction of electron tracks determines a vector that points back to the source of the neutrinos. In the case of solar neutrino scattering events, this is the Sun. Radiochemical detectors cannot determine the direction of neutrinos, and their origin must be inferred from calculations of neutrino fluxes and energies from possible candidates. Thirdly, the energy distribution of the recoil electrons reflects the energy distribution of the incident neutrinos.

\subsubsection{Inelastic Scattering}
If the neutrinos have enough energy, they can produce resonance states much heavier than the proton or neutron, which decay into a nucleon and one or more pions. This type of interaction is known as inelastic scattering. A typical example is:
\begin{eqnarray*}
\bar{\nu_{e}} + p \rightarrow n + e^{+} + \pi^{0}.
\end{eqnarray*}
Events of this particular type are the main source of background in proton decay searches. The observed products from this interaction are the same as those expected from some proton decay channels, for example, $p \rightarrow e^{+} \pi^{0}$. The most energetic solar neutrinos (from \Bor\ beta-decay) have insufficient energy to participate in this type of interaction; their maximum energy is 14~MeV.

\subsubsection{Quasi-Elastic Neutrino-Nucleon Scattering}
The Feynman diagram for quasi-elastic neutrino-nucleon scattering is shown in figure~\ref{capture} for the case with electron neutrinos. If, instead of an electron neutrino, a muon neutrino interacts then a muon will be produced. Therefore it is possible to infer the presence of an electron neutrino by observing the creation of an electron, and a muon neutrino from the muon produced.

\begin{figure}[!htbp]
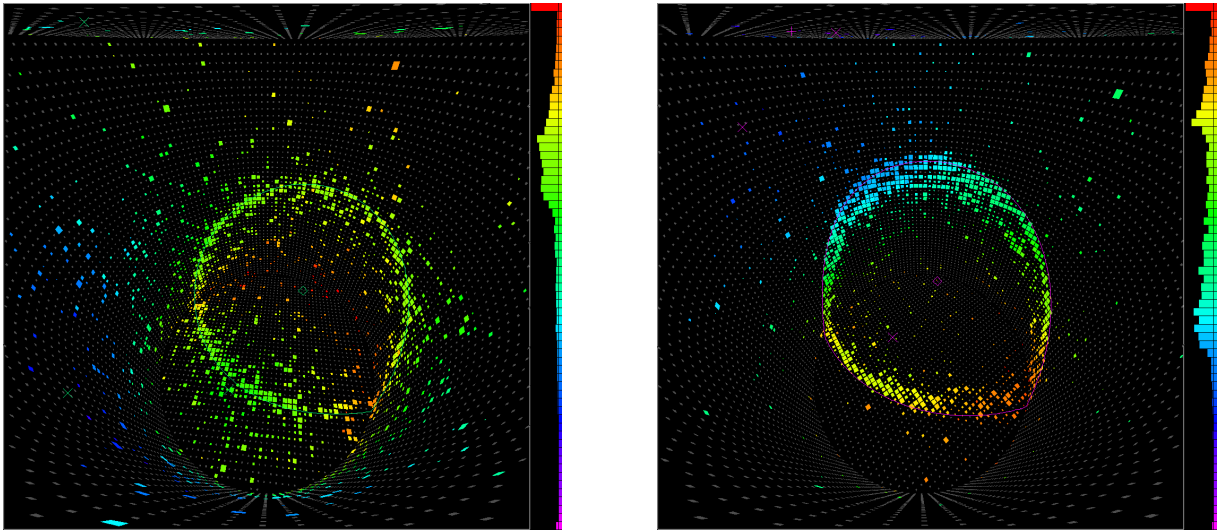
 
\setlength{\unitlength}{1cm}
\epsfig{file=eevent.eps,height=7cm} % 7cm
\hspace{1.0cm}
\epsfig{file=muevent.eps,height=7cm} % 7cm
\caption{\small{An electron neutrino event (left) and a muon neutrino event (right) at Super-Kamiokande. (From~\cite{Tomba}).}}
\label{rings} 
\end{figure}

The \Cerenkov\ PMT hit patterns for electron-type and muon-type events are different. Muons can leave the detector with most of their energy intact, and produce well-defined \Cerenkov\ rings on the PMT array. The PMT hit patterns for electron-type events are different because the energetic electrons produce an electromagnetic shower of secondary particles by bremsstrahlung and pair production. Each of these secondary particles produce tracks that are slightly spread in direction, and each track produces its own \Cerenkov\ cone. The multiple \Cerenkov\ cones produce a diffuse \Cerenkov\ ring in the PMT array. For each event, the triggering and data acquisition system is able to identify the neutrino type on the basis of whether the event was showering (electron-type) or non-showering (muon-type). Examples of each event type are shown in figure~\ref{rings}.

%%%%%%%%%%%%%%%%%%%%%%%%%%%%%%%%%%%%%%%%%%%%%%%%%%%%%%%%%%%%%%%%%%%%%%%%%%%%%%
\subsection{Supernova 1987A}
The work to upgrade the Kamiokande detector to make it sensitive to solar neutrinos was completed by late 1986, just in time to observe the neutrinos emitted 170000 years earlier by a supernova in the Large Magellanic Cloud. Fortunately supernova neutrinos and solar neutrinos have similar energies (about 10~MeV). The supernova, named SN1987A, created a massive pulse of neutrinos that were detected simultaneously by the Kamiokande~II and IMB detectors, and also by the Baksan liquid scintillation neutrino telescope, located in the Caucasus Mountains of Russia. This was the first ever detection of neutrinos from a source outside the solar system.

The first events were detected on 23~February~1987 at about 0736~UTC. Kamiokande~II detected twelve neutrino events in total that could be attributed to SN1987A~\cite{JNBbook}. The time interval, recorded by Kamiokande~II, between the first event and the last was 12.439~s, and the recoil electrons were found to have energies between 6.3$\pm$1.7~MeV and 35.4$\pm$8.0~MeV. IMB detected eight neutrino events from SN1987A~\cite{JNBbook}. IMB recorded the time interval between the first and last event to be 5.582~s. The recoil electron energies were in the range 19$\pm$5~MeV to 38$\pm$7~MeV. The indicated errors are one standard deviation uncertainties. No events above the solar neutrino background and attributable to SN1987A were detected by the Homestake \Chl\ detector~\cite{JNBbook}.

Observation of neutrinos from SN1987A led to important inferences about neutrino properties. New limits were obtained for neutrino mass, charge, magnetic moment, decay rate, limiting velocity, and total number of flavours~\cite{JNBbook}. For example, if neutrinos have a finite mass, and if they are emitted simultaneously, one would expect the higher energy neutrinos from a supernova to arrive before the slower, lower energy, neutrinos. Hence, a non-zero neutrino mass results in a spread of arrival times. An upper limit of 16~eV for the mass of the electron neutrino has been derived statistically by fitting Monte Carlo simulations with the observed SN1987A data~\cite{JNBbook}. %The OMNIS (Observatory for Multiflavor NeutrInos from Supernovae) collaboration have proposed to build a lead/iron detector in New Mexico that will use precision time-of-arrival measurements of neutrinos from supernovae to determine neutrino mass differences and oscillation parameters.
By a similar argument an upper limit can be set for the electric charge of an electron neutrino. If electron neutrinos have a finite charge they will be deflected by the galactic magnetic field. One would expect higher energy neutrinos to move in a straighter path, and therefore arrive sooner, than lower energy neutrinos. It has also been noted that the fact that neutrinos have indeed been observed from SN1987A implies a lower limit of 10$^{5}$ years for the lifetime of a 10~MeV electron antineutrino. Moreover, the approximate equality of the arrival times of the first photons and neutrinos from SN1987A implies that the speed of photons and neutrinos cannot differ by more than 1 part in 10$^{8}$.~\cite{Stodolsky,JNBbook}

\subsection{Solar Neutrino Results}
Preliminary results from the Kamiokande~II detector reported the observed solar neutrino flux to be 45\percent\ of that predicted by theory~\cite{JNBbook,JNB3,Kamioka}. The experiment ran for almost a decade, and data-taking officially ended in April 1996. The most recent theoretical prediction for the detection rate at Kamiokande~II is 5.15$\times$10$^{6}$~cm$^{-2}$s$^{-1}$. Final results from the Kamiokande~II detector show that it observed a rate of 2.80$\times$10$^{6}$~cm$^{-2}$s$^{-1}$; 54\percent\ of that predicted by theory\footnote{In the literature, detection rates for water \Cerenkov\ detectors are often expressed in units of 10$^{6}$~cm$^{-2}$s$^{-1}$. Results are often expressed in terms of a ratio to the expected event rate.}~\cite{nunews,BPtwoK}. Kamiokande~II established experimentally that these neutrinos were coming directly from the Sun: the recoil electrons were scattered predominantly in the direction of the Sun-Earth vector. In addition, the observed neutrino energies are consistent with the range of energies (0 -- 15~MeV) expected from standard solar model predictions.

The marked discrepancy between the predicted and measured neutrino fluxes observed by the Kamiokande~II detector confirmed the earlier result observed by the Homestake \Chl\ detector. This was of great importance because observational results for the previous two decades had come from one experiment, using a different detection method.

\subsubsection{Incompatibility of \Chl\ and Water \Cerenkov\ Experiments}
The same \Bor\ neutrinos that dominate the Homestake \Chl\ capture rate also determine the water \Cerenkov\ event rate. Therefore, one can calculate the Homestake \Chl\ capture rate that is produced by the \Bor\ neutrinos observed in a water \Cerenkov\ experiment. This capture rate is predicted to be 2.78$\pm$0.10~SNU, which exceeds (by one standard deviation) the \emph{total} observed capture rate of 2.56$\pm$0.23~SNU. This means that the calculated net contribution to the capture rate in the Homestake \Chl\ detector from all other neutrino sources ($pep$, \Ber\, and CNO neutrinos) is negative! This makes no sense unless something happens to the \Bor\ neutrinos in transit that changes their energy spectrum. If the Standard Model is correct, then the shape of the energy spectrum should not deviate appreciably from that determined by laboratory experiments. Therefore, new physics must be responsible for the discrepancy. It is important to note that no solar model considerations are required to reach this conclusion. The incompatibility of the Homestake \Chl\ and water \Cerenkov\ experiments is the \emph{second} solar neutrino problem~\cite{JNB2,JNBEAA}.

\subsection{Atmospheric Neutrinos}
Kamiokande~II (and IMB) initially began studying atmospheric neutrinos because they were a significant source of background in proton decay searches. Atmospheric neutrinos could mimic certain types of proton decay events. Atmospheric neutrinos are produced in hadronic showers induced by primary cosmic rays (typically protons) striking nuclei in the upper atmosphere. The cosmic rays produce pions (mostly) and kaons that decay to muons and muon neutrinos (and their anti-particles). The muons then decay, producing electrons, muon neutrinos, and electron neutrinos (and their anti-particles also). The production process is summarised by the following chain of reactions:
\begin{center}
\begin{tabular}{l l l l}
$p\;+\;air~\rightarrow$ & $\pi^{\pm}(K^{\pm})$ & $+$ & $X$\\
&$\pi^{\pm}(K^{\pm})$ & $\rightarrow$ & $\mu^{\pm}\;+\;\nu_{\mu}(\bar{{\nu_{\mu}}})$\\
&&&$\mu^{\pm}\;\rightarrow\;e^{\pm}\;+\;\nu_{e}(\bar{\nu_{e}})\;+\;\bar{\nu_{\mu}}(\nu_{\mu})$.
\end{tabular}
\end{center}
Naively one would expect the following ratio by counting the decay neutrinos:
\begin{eqnarray*}
\frac{\nu_{\mu} + \bar{\nu_{\mu}}}{\nu_{e} + \bar{\nu_{e}}} \approx \frac{2}{1}
\end{eqnarray*}
However the ratio depends on the energies of the neutrinos. A thorough study must take into account the differences of the lifetimes and spectra of the pions, kaons, and muons~\cite{Akhmedov}.

Unfortunately, the ratio of showering (electron-type) and non-showering (muon-type) events in a water \Cerenkov\ detector will not be an accurate reflection of the ratio of muon neutrinos to electron neutrinos being produced in the atmosphere. The response of a detector depends on several factors, such as neutrino type and energy; cross sections for neutrino interactions will depend both. Furthermore, different detectors have different responses. Therefore, the ratio of neutrino types is a meaningful statistic only if compared against the expected ratio for the detector. A useful statistic is the double ratio:
\begin{eqnarray*}
R = \frac{\left(\frac{\nu_{\mu} + \bar{\nu_{\mu}}}{\nu_{e} + \bar{\nu_{e}}}\right)_{Observed}}{\left(\frac{\nu_{\mu} + \bar{\nu_{\mu}}}{\nu_{e} + \bar{\nu_{e}}}\right)_{Monte\;Carlo}}.
\end{eqnarray*}
The ratio in the denominator, as the subscript suggests, is determined from Monte Carlo simulations. If the measured result agrees with the result from the Monte Carlo simulation, the ratio will be equal to unity.  Kamiokande~II (and IMB) found the double ratio $R$ to be about 0.6~\cite{nunews,Akhmedov,Langacker}. This result has been confirmed by the Super-Kamiokande, Soudan~2, and MACRO (Monopole, Astrophysics and Cosmic Ray Observatory) experiments~\cite{nunews,Akhmedov,Langacker}. It is worth noting that the latter two experiments use different detection techniques than those employed at Kamiokande~II and IMB. The Soudan~2 detector is an iron calorimeter~\cite{Soudan}. The MACRO detector uses tanks of liquid scintillator, planes of streamer tubes, and plates of track etch material~\cite{MACRO}. This persistent discrepancy between the observed and predicted neutrino fluxes has become known as the \emph{atmospheric neutrino anomaly}.

\section{Gallium Experiments}
The reaction that is used to detect solar neutrinos in gallium experiments is:
\begin{eqnarray*}
\nu_{e} + {^{71}Ga} \rightarrow e^{-} + {^{71}Ge}.
\end{eqnarray*}
The threshold energy of this reaction is 0.2332~MeV~\cite{JNBbook}. The low threshold energy is of critical importance to gallium experiments, because it is this that makes possible the detection of the $pp$ solar neutrinos that \Chl\ and water \Cerenkov\ experiments are insensitive to (see figure~\ref{snspec}). These neutrinos are important because they dominate the total flux of of neutrinos from the Sun. The \Bor\ neutrino flux is tiny in comparison, about 10$^{-4}$ of the $pp$ neutrino flux. The flux of the $pp$ neutrinos is predicted to an accuracy of 1\percent, and has the lowest uncertainty of all the predicted neutrino fluxes~\cite{JNBEAA}. The theoretical uncertainty in the predicted \Bor\ neutrino flux is about 19\percent~\cite{JNBEAA}. The chemical processing differs from \Chl\ radiochemical experiments, but the underlying physical principles are the same. The capture rate predicted by the standard solar model for a \Gal\ detector is 128$^{+9}_{-7}$~SNU (1$\sigma$ errors)~\cite{BPtwoK}.

\subsection{GALLEX}
The GALLEX (GALLium EXperiment) detector was located at the Gran Sasso Underground Laboratory in Italy, and consisted 30~tons of gallium in an aqueous solution of gallium chloride and hydrochloric acid in a single vessel.

A known number of non-radioactive germanium atoms was added to the solution at the beginning of each run, so that the extraction efficiency could be determined experimentally at the end of each run. The neutrino-induced \Ger\ atoms forms GeCl$_{4}$, which was purged from the solution at the end of each run by bubbling nitrogen gas through the tank. The nitrogen was then passed through a gas scrubber where the GeCl$_{4}$ was absorbed in water. The GeCl$_{4}$ was converted to GeH$_{4}$ through a series of chemical processes. The GeH$_{4}$ was then loaded into a small proportional counter, and the number of \Ger\ atoms determined by observing their radioactive decay.

GALLEX took data between 1991 and 1997. The observed capture rate, calculated from data combined from all 65 runs, was 77.5$^{+7.6}_{-7.8}$~SNU (1$\sigma$ errors)~\cite{GNOpaper}, 60\percent\ of that predicted by theory.

\subsection{GNO}
GNO (Gallium Neutrino Observatory) is the successor project of GALLEX at Gran Sasso. GNO started taking data in April 1998. Over the next few years the gallium mass will be increased from the present 30~tons up to 100~tons. The first phase of the experiment, GNO30, is just a prolongation of GALLEX with upgraded electronics and proportional counters. The second phase, GNO66, will involve increasing the target mass to the maximum 66~tons that can be accommodated in the two available tanks. The final addition of mass in the third phase of the experiment, GNO100, will possibly be in metallic form.

The first results from GNO for the measuring period 20~May~1998 to 12~January~2000 were presented at the Neutrino~2000 conference in Sudbury, Canada. GNO reported a \Gal\ production rate equivalent to 65.8$^{+10.7}_{-10.2}$~SNU (1$\sigma$ errors)~\cite{GNOpaper}. If the initial data from GNO and the data from all 65 runs in the GALLEX experiment are combined the resultant rate is 74.1$^{+6.7}_{-6.8}$~SNU (1$\sigma$ errors) \cite{GNOpaper}, 58\percent\ of that predicted by theory.

\subsection{SAGE}
SAGE (Soviet-American Gallium Experiment) was located in an underground chamber at the Baksan Neutrino Observatory in the Caucasus Mountains, Russia, and used a 60~ton metal gallium target contained in ten identical reactor vessels. The germanium extraction procedure involves melting the gallium metal (at about 30~\degC) and mixing it in dilute hydrochloric acid. The \Ger\ is removed from the solution using a procedure similar to that used in the GALLEX experiment.

SAGE started taking data in 1990 and is still running. So far, SAGE has observed a capture rate of 75.4$^{+7.8}_{-7.4}$~SNU (1$\sigma$ errors) \cite{BPtwoK}, 59\percent\ of that predicted by theory.

\subsection{The Gallium Problem}
The results reported from the GALLEX and the SAGE experiments are consistent with each other. The average of the SAGE, GNO, and GALLEX results is 74.7$\pm$5.0~SNU, more than 6$\sigma$ away from the theoretical value~\cite{BPtwoK}.

The theoretical rate calculated for the $pp$ and $pep$ neutrinos fully accounts for the observed rate in the gallium experiments. However, the \Bor\ neutrinos observed in the water \Cerenkov\ experiments must also contribute to the gallium event rate. The theoretical \Bor\ partial event rate can be normalised to the rate observed in a water \Cerenkov\ experiment. When this partial event rate is added to sum of the event rates for the $pp$ and $pep$ neutrinos (72.5~SNU~\cite{BPtwoK}) -- to account for the contribution from the \Bor\ neutrinos -- it is found there is no room for the additional 34.2~SNU~\cite{BPtwoK} expected from \Ber\ neutrinos, given the measured rates in the gallium experiments. The \Ber\ neutrinos appear to be missing. To account for this one would have to set the rates of the all reactions that produce \Ber\ in the Sun to zero. However, \Bor\ is produced by proton capture on \Ber, and \Bor\ neutrinos are observed! This is the \emph{third} solar neutrino problem.

\section{Super-Kamiokande}
Super-Kamiokande is a water \Cerenkov\ detector located 1~km underground in the Kamioka zinc mine in the Japanese Alps. The detector consists of a stainless steel cylindrical tank, 41.4~m in height by 39.3~m in diameter~\cite{Kamioka}, filled with 50,000 tons of pure water. The water tank consists of an inner detector and an outer detector. The inner detector is instrumented by 11,146 50~cm diameter PMTs facing inward~\cite{SKichep}. The outer detector surrounds the inner detector and is instrumented by 1885 20~cm diameter PMTs facing outward~\cite{SKichep}. Only the inner 22,500~tons of water, located more than 2~m inwards from the inner detector wall, is used for solar neutrino experiments and proton decay searches~\cite{SKichep,SKosc}. The outer volume of water is used to tag incoming high-energy cosmic ray muons and to attenuate background radiation from outside the tank. The trigger of the detector accepts events above an energy threshold of 5.7~MeV~\cite{SKichep}. A new trigger scheme installed in April 1997 should enable data to be taken at a lower threshold (4.7~MeV) in future~\cite{SKichep}. Figure \ref{snspec} shows that the solar neutrino event rate in the Super-Kamiokande detector will be dominated by the \Bor\ neutrinos. Super-Kamiokande has been taking data since 1 April 1996~\cite{Kamioka}.
\begin{figure}[!htbp]
\begin{center}
\mbox{\epsfig{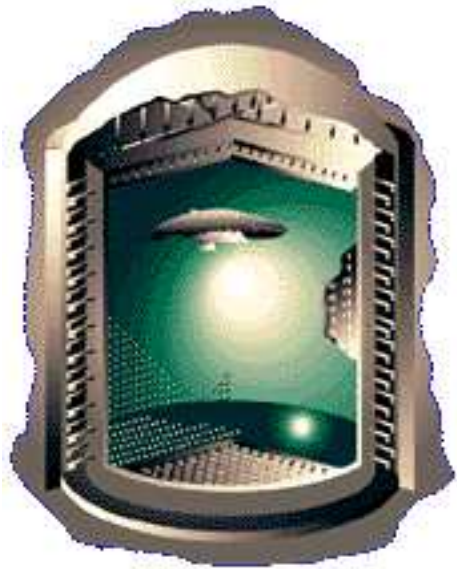}}
\caption{\small{The Super-Kamiokande detector. The 60~ft humpback whale is shown for size comparison and is not part of the experiment.}}
\label{superk}
\end{center}
\end{figure}
\subsection{Results from Super-Kamiokande}
The Super-Kamiokande experiment reported the observed solar neutrino flux to be 47\percent\ of that predicted by theory~\cite{SKichep}. Furthermore, Super-Kamiokande has measured a value for the double ratio $R$ that is consistent with that measured by Kamiokande~II~\cite{SKichep}. Super-Kamiokande recently found convincing evidence for oscillations of atmospheric neutrinos~\cite{SKosc}. This implies that neutrinos have mass. This important discovery will be discussed in greater detail later (see section 9).
\section{SNO}
SNO (Sudbury Neutrino Observatory) is a water \Cerenkov\ detector, but unlike Kamio-kande~II or Super-Kamiokande, it uses heavy water. The SNO detector is located 2~km underground in the Creighton nickel mine near Sudbury, Ontario, Canada. SNO will be able to measure the total flux of neutrinos, of any flavour, from the Sun~\cite{JNBbook}. SNO \mbox{detects} solar neutrinos in three different channels; elastic neutrino-electron scattering (which can proceed through both CC and NC channels -- see figures~\ref{scat1}, \ref{scat2}, \ref{scat3}, and \ref{scat4}), and the disintegration of deuterium~\cite{SNO}:
\begin{eqnarray*}
\nu_{e} + d &\rightarrow& p + p + e^{-}~~~(CC)\\
\nu_{l} + d &\rightarrow& p + n + \nu_{l}~~~(NC)
\end{eqnarray*}
The former reaction involves the charged current (CC) weak interaction, and can only be initiated by electron neutrinos. The latter reaction involves the neutral current (NC) weak interaction, and occurs with equal probability for all \emph{active} (\nue, \numu, \nutau) neutrinos. %\nue\goesto\numu\ or \nue\goesto\nutau\ oscillations would not change the NC event rate, but would deplete the solar \nue\ flux, reducing the CC event rate.
Therefore the CC/NC ratio is a sensitive probe of neutrino oscillations. SNO started taking data at the end of 1999, but has not published any results yet other than the confirmation that solar neutrinos have been observed in the detector~\cite{SNO}.
\section{Evidence for New Physics}
\subsection{Three Solar Neutrino Problems}
\begin{figure}[!htbp]
\begin{center}
%\vspace{-1.5cm}
\mbox{\epsfig{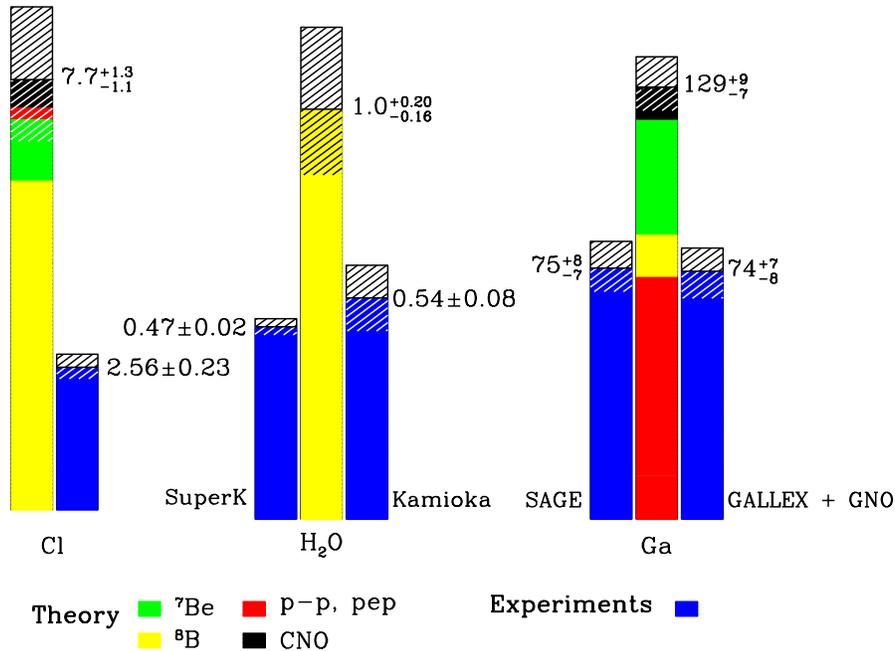}}
%\vspace{-0.5cm}
\caption{\small{Comparison between the predicted and observed fluxes in the \Chl, water \Cerenkov, and \Gal\ experiments. In~\cite{BPtwoK}, Bahcall, Basu, and Pinsonneault have used the most recently determined value of the solar luminosity to calculate neutrino capture rates. This value is 0.2\percent\ smaller than the value used to calculate the capture rates shown in this figure. As a result, the predicted \Bor, \Ber, and absolute neutrino fluxes shown are 2\percent, 1\percent, and 1~SNU greater, respectively. The water \Cerenkov\ result is expressed in terms of a ratio to the expected event rate. (From~\cite{JNBhomepage}).}}
\label{exprates}
\end{center}
\end{figure}

Figure \ref{exprates} illustrates the three solar neutrino problems:
\begin{enumerate}
\item The discrepancy between the observed solar neutrino flux and that predicted by the standard solar model.
\item The incompatibility of the \Chl\ and water \Cerenkov\ experiments.
\item The absence of \Ber\ neutrinos implied by the rate in the gallium experiments.
\end{enumerate}
\newpage
\noindent There are three possible solutions to the solar neutrino problem:
\begin{enumerate}
\item The standard solar model is incorrect.
\item The gallium experiments, plus either the \Chl\ or water \Cerenkov\ detectors, have yielded misleading results.
\item Something happens to the neutrinos in transit from the Sun to the detector that is not explained by the Standard Model.
\end{enumerate}
Although it is not \emph{impossible} that the standard solar model is at fault, it seems very unlikely. There is much experimental evidence to support the claim that the standard solar model is an accurate description of the Sun. It has been very successful in explaining the physical parameters and characteristics of the Sun. Moreover, helioseismological measurements have demonstrated that the sound speeds predicted by the standard solar model are extraordinarily accurate. It is hard \emph{not} to be convinced of the standard solar model's validity. However, there are a number of solutions to the three solar neutrino problems that involve non-standard solar models. All the solutions involve assumptions that are generally regarded as unlikely or impossible. All can therefore be rejected by applying the principle of parsimony (otherwise known as Occam's Razor).

If the solar neutrino problem is due to experiment-related errors, one must conclude that at least three of the five experiments shown in figure~\ref{exprates} have yielded erroneous results. This seems to be very unlikely; each detector has been calibrated and tested thoroughly, and the shortfall in the predicted absolute neutrino flux is observed by \emph{all} the detectors.

The final option is a particle physics solution. An obvious solution is that physics beyond the Standard Model produces some change in the neutrinos on their journey from the Sun to the Earth so that fewer are observed in the experiments.

\subsection{The Atmospheric Neutrino Anomaly}
The low value for the double ratio $R$ implies that the flavour content of atmospheric neutrinos is changed after they are created. One might guess that the mechanism responsible for the atmospheric neutrino anomaly is somehow related to the mechanism responsible for the solar neutrino problem. Therefore it is desirable to seek a solution that will explain both phenomena. 
%%%%%%%%%%%%%%%%%%%%%%%%%%%%%%%%%%%%%%%%%%%%%%%%%%%%%%%%%%%%%%%%%%%%%%%%%%%%%%
\section{Theory}
\subsection{Particle Physics Solutions}
There are a number of different particle physics solutions to the solar neutrino problem. These include neutrino decay, resonant spin conversion, WIMP effects, and neutrino oscillations.

The first of these solutions, neutrino decay, is disfavoured because it requires that electron neutrinos have a lifetime on the order of 8 minutes, the transit time for the distance between the Sun and the Earth. The observation of electron antineutrinos from SN1987A requires that their lifetime is at least 10$^{5}$ years, hence it is extremely unlikely that the solar neutrino problem can be explained by neutrino decay~\cite{JNBbook}.

The second solution listed, resonant spin conversion, requires that the electron neutrino has a large magnetic moment. If this condition is satisfied, passage through the solar magnetic field may flip the spin of neutrinos. The spin flip would change a left-handed neutrino into a right-handed neutrino~\cite{JNBbook}. It is known experimentally that antineutrinos, which are right-handed, cannot participate in reactions caused by neutrinos, which are left-handed. This is called \emph{chiral prohibition}~\cite{Akhmedov}. The weak interactions that underlie the neutrino capture reactions in the \Chl\ and \Gal\ experiments are chiral. Right-handed neutrinos have the \emph{wrong chirality} and are not detectable in these experiments. If neutrinos are massive, then neutrinos that have been spin-flipped into a right-handed state will have a very small left-handed component, so that neutrino capture on \Chl\ or \Gal\ is possible -- but extremely unlikely~\cite{Akhmedov}. The spin flip mechanism is disfavoured because it requires a neutrino magnetic moment many orders of magnitude greater than typical theoretical values calculated with conventional models~\cite{JNBbook}.

Weakly interacting massive particles (WIMPs) have been suggested as a solution to the solar neutrino problem and the dark matter problem. It is believed that WIMPs could facilitate energy transport in the Sun and thereby lower the temperature gradient, and therefore decrease production of \Bor\ neutrinos, in the solar core~\cite{JNBbook}. This hypothesis is also disfavoured because it requires a new particle with unusual characteristics.

The last of the mechanisms listed, neutrino oscillations, requires the least number of assumptions in the formulation of a solution to both the solar neutrino problem and the atmospheric neutrino anomaly. In accordance with Occam's Razor, this is the preferred explanation. The idea that neutrino oscillations may explain the solar neutrino problem was first introduced by Bruno Pontecorvo~\cite{Bruno}.

\subsection{Neutrino Oscillations}
The Standard Model does not explain the observed masses for quarks and leptons. Neutrinos are assumed to massless in the Standard Model, but there is no fundamental reason why this should be~\cite{PPbook}. If neutrinos have non-zero masses their flavour eigenstates do not have to coincide with their mass eigenstates. Instead, flavour eigenstates are linear superpositions of mass eigenstates~\cite{PPbook,PDG}. Consequently, a new phenomenon can occur: \emph{neutrino mixing}. This is directly analogous to quark mixing. The flavour eigenstates $\nu_{l}~(l=e, \mu, \tau)$ are given by
\begin{eqnarray*}
\left|{\nu_{l}}\right> = \sum_{m} U_{ln} \left|{\nu_{n}}\right>
\end{eqnarray*}
where $\nu_{n}~(n=1, 2, 3)$ are the mass eigenstates and the coefficients $U_{ln}$ form a 3 $\times$ 3 \mbox{unitary} matrix, $U$, known as the leptonic mixing matrix. The matrix $U$ is also known as \mbox{\emph{Maki-Nakagawa-Sakata (MNS) matrix}}~\cite{PDG}. It is the leptonic analogue of the Cabibbo-Kobayashi-Maskawa (CKM) matrix~\cite{Akhmedov}. More than three mass eigenstates may exist, but usually it is assumed that no more than three make significant contributions to the flavour eigenstates~\cite{PDG}.

In general, $U$ does not have a simple form. For the case with three flavours, the coefficients $U_{ln}$ depend on \emph{three} mixing angles~\cite{Akhmedov}. The important aspects of neutrino oscillations can be understood by studying the simple case with just two neutrino flavours, \nue\ and \numu\ for example. Most neutrino oscillation analyses only consider two-flavour mixing scenarios. This is a good approximation if the mass scales are quite different, because then the oscillation phenomena tend to decouple, and each transition can therefore be described by a two-flavour mixing equation~\cite{RRichep}. For two-flavour \nue--\numu\ mixing, the lepton mixing matrix $U$ can be written
\begin{center}
\begin{math}
U = \left(\begin{array}{r r}
\cos{\theta} & \sin{\theta} \cr
{-\sin{\theta}} & \cos{\theta} \cr
\end{array}\right)
\end{math}
\end{center}
where $\theta$ is the \nue--\numu\ mixing angle. In this case, a flavour eigenstate can spontaneously change into an eigenstate of another flavour as the neutrino propagates in space. This is the phenomenon known as \emph{neutrino oscillation}. For example, consider an electron neutrino with momentum $p$ at time $t = 0$. The initial state is
\begin{eqnarray*}
\left|{\nu_e}\right> = \cos{\theta}{\left|{\nu_1}\right>} + \sin{\theta}{\left|{\nu_2}\right>}
\end{eqnarray*}
but at time $t$ this will become
\begin{eqnarray*}
\left|{\nu_e}\right>_t = {e^{-i{E_1}t}}\cos{\theta}{\left|{\nu_{1}}\right>} + {e^{-i{E_2}t}}\sin{\theta}{\left|{\nu_{2}}\right>}
\end{eqnarray*}
where $E_1$ and $E_2$ are the energies of the two mass eigenstates with the same momentum $p$. Note that units are adopted in which $\hbar = c = 1$. The probability amplitude for a \nue\ being observed at time $t$ is
\begin{eqnarray*} 
\left<{{\nu_e}|{\nu_e}}\right>_t = {e^{-i{E_1}t}}\cos^{2}{\theta} + {e^{-i{E_2}t}}\sin^{2}{\theta}.
\end{eqnarray*}
Therefore, the probability that at time $t$ the \nue\ will retain its original flavour is
\begin{eqnarray*}
P({\nu_e}\rightarrow{\nu_e}) = \left|\left<{{\nu_e}|{\nu_e}}\right>_t\right|^{2} = 1 - \sin^{2}{2\theta}\sin^{2}\left({\frac{1}{2}}\left({E_2} - {E_1}\right)t\right).
\end{eqnarray*}
The probability that at time $t$ the flavour of the neutrino is \numu\ is
\begin{eqnarray*}
P({\nu_e}\rightarrow{\nu_{\mu}}) = 1 - P({\nu_e}\rightarrow{\nu_e}) = \sin^{2}{2\theta}\sin^{2}\left({\frac{1}{2}}\left({E_2} - {E_1}\right)t\right).
\end{eqnarray*}
For relativistic neutrinos with momentum $p$,
\begin{eqnarray*}
E_{m} = \sqrt{p^{2} + m^{2}_{n}} \approx p + \frac{m^{2}_{n}}{2p} \approx p + \frac{m^{2}_{n}}{2E},
\end{eqnarray*}
the energy difference is
\begin{eqnarray*}
{E_2} - {E_1} = \frac{m^{2}_{2} - m^{2}_{1}}{2E} = \frac{{\Delta}m^2}{2E}
\end{eqnarray*}
where
\begin{eqnarray*}
{\Delta}m^2 = m^2_2 - m^2_1,
\end{eqnarray*}
and $m_1$ and $m_2$ are the masses of the mass eigenstates. It is convenient to write the transition probabilities in terms of the distance $L$ travelled by the neutrinos. For relativistic neutrinos $L \approx t$. Therefore the transition probability $P($\nue\goesto\numu$)$ may be written
\begin{eqnarray*}
P({\nu_e}\rightarrow{\nu_{\mu}}) = \sin^{2}{2\theta}\sin^{2}\left({\pi\frac{L}{L_{osc}}}\right)
\end{eqnarray*}
where the oscillation length $L_{osc}$ is defined as
\begin{eqnarray*}
L_{osc} = \frac{4{\pi}E}{{\Delta}m^2} \approx 2.48~m\frac{E~(MeV)}{{\Delta}m^{2}~(eV^2)} \approx 2.48~km\frac{E~(GeV)}{{\Delta}m^{2}~(eV^2)}.
\end{eqnarray*}
Therefore
\begin{eqnarray*}
P({\nu_e}\rightarrow{\nu_{\mu}}) = sin^{2}{2\theta}\sin^{2}\left({1.27{{\Delta}m^{2}}\frac{L}{E}}\right)
\end{eqnarray*}
where $L$ is in $m$ and $E$ is in MeV, or $L$ is in $km$ and $E$ is in GeV. Clearly the oscillations vanish if the mixing angle ${\theta}$ is zero, or if the neutrinos have equal masses, or are both massless. The mechanism just described is known as the \emph{vacuum oscillation} (VO) solution, and was first discussed by Vladimir Gribov and Bruno Pontecorvo~\cite{Bruno}. Vacuum oscillations are believed to be an unlikely (but not excluded) solution to the solar neutrino problem because fine tuning of neutrino parameters is required, and the necessary mixing angles must be much larger than the known quark mixing angles~\cite{JNBbook}. The VO solution is also known as the ``Just So'' hypothesis, because it assumes that the position of the Earth from the Sun coincides with an oscillation maximum~\cite{RRichep}.

\subsubsection{The MSW Effect}
It is thought that neutrino interactions in matter can enhance oscillations. This process is known as the \emph{MSW effect}. The MSW effect is currently the most popular solution to the solar neutrino problem because, unlike the vacuum oscillation solutions, no fine tuning of neutrino parameters is required; the mixing angles and mass differences can vary by orders of magnitude~\cite{JNBbook,Smirnov,Wolfenstein}. The MSW effect takes its name from Stanislav Mikheyev, Alexei Smirnov, and Lincoln Wolfenstein.

Electron neutrinos created in fusion reactions in the Sun may be transformed into neutrinos of another flavour via the MSW effect. The MSW conversion results from interactions between neutrinos and electrons in the Sun~\cite{JNBbook}. Therefore neutrinos that were created with flavour \nue\ can be transformed into neutrinos of other flavours that are more difficult to detect. Muon neutrinos and tau neutrinos are invisible in the \Chl\ and \Gal\ experiments. Muon neutrinos and tau neutrinos can scatter off electrons in water \Cerenkov\ detectors via the neutral current interaction, but the cross section for this reaction is about a factor of six smaller than the cross section for the charged current channel~\cite{nunews}. Hence the solar neutrino problem is explained as the result of conversions in the Sun of electron neutrinos to neutrinos of other flavours that are more difficult to detect; these are either muon neutrinos, tau neutrinos, or perhaps sterile neutrinos~\cite{JNBbook}. The sterile neutrino, denoted $\nu_s$, is a right-handed~\cite{Langacker} isosinglet partner to the ``standard light neutrinos'', and does does not participate in weak interactions~\cite{PDG}. %The SNO experiment will be able to test the hypothesis that \nue\goesto\nuster\ oscillations are responsible for the solar neutrino problem.

It is believed that the MSW effect can also occur in the Earth. For example, solar neutrinos that shine \emph{up} on us during the night have travelled some distance through the Earth on their way to the detector. Muon neutrinos and tau neutrinos that were produced by \emph{matter-enhanced} neutrino oscillations in the Sun can be be \emph{reconverted} into electron neutrinos within the Earth~\cite{nunews,JNBbook}. This leads to a day/night difference due to the \emph{regeneration} of electron neutrinos in the Earth via the MSW effect. This difference can, in principle, be observed in real-time experiments, such as Super-Kamiokande~\cite{nunews}. The Sun should appear brighter in neutrinos at night. Super-Kamiokande found an excess of the night-time flux at about 1.3$\sigma$, but this is not statistically significant~\cite{JNB5,nunews,SKichep}.

\begin{figure}[!htbp]
\begin{center}
\epsfig{file=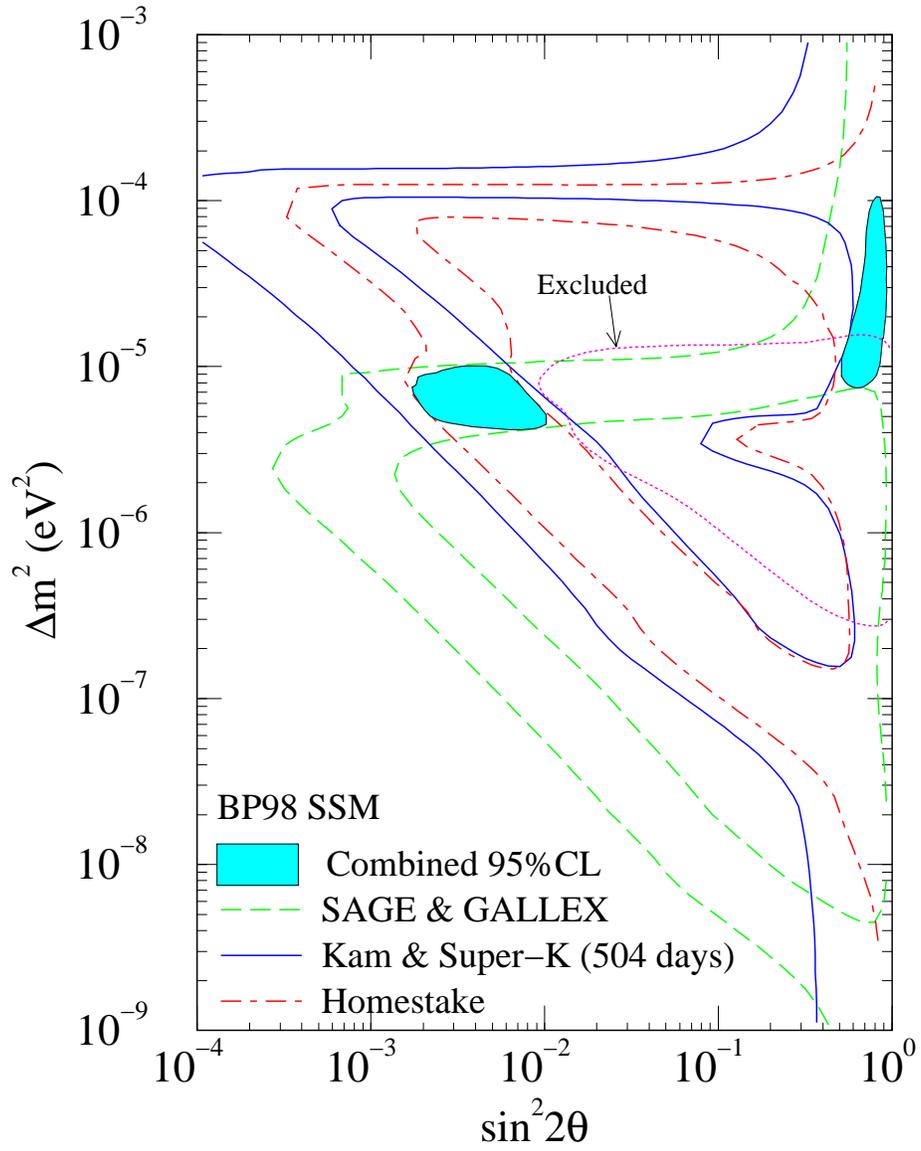,width=12cm}
\caption{\small{Allowed MSW solutions (shaded areas) to the solar neutrino problem. (From~\cite{Langacker}).}}
\label{solarMSW}
\end{center}
\end{figure}

There are two possible solutions of the solar neutrino problem that make use of the MSW effect; either \mbox{${\Delta}m^2 \sim$ 10$^{-5}$~eV$^2$} and \mbox{$\sin^2{2\theta} \sim$ 10$^{-2}$} or \mbox{${\Delta}m^2 \sim$ 10$^{-5}$~eV$^2$} and  \mbox{$\sin^2{2\theta} \sim$ 0.6}~\cite{JNB2}, where $\theta$ is the mixing angle and ${\Delta}m^2$ is the difference in the squared masses of the mass eigenstates. Figure~\ref{solarMSW} shows these two solutions on an \emph{MS diagram}. MS diagrams are discussed in the next section. 

\subsubsection{MS Diagrams}
\begin{figure}[!htbp]
\begin{center}
\epsfig{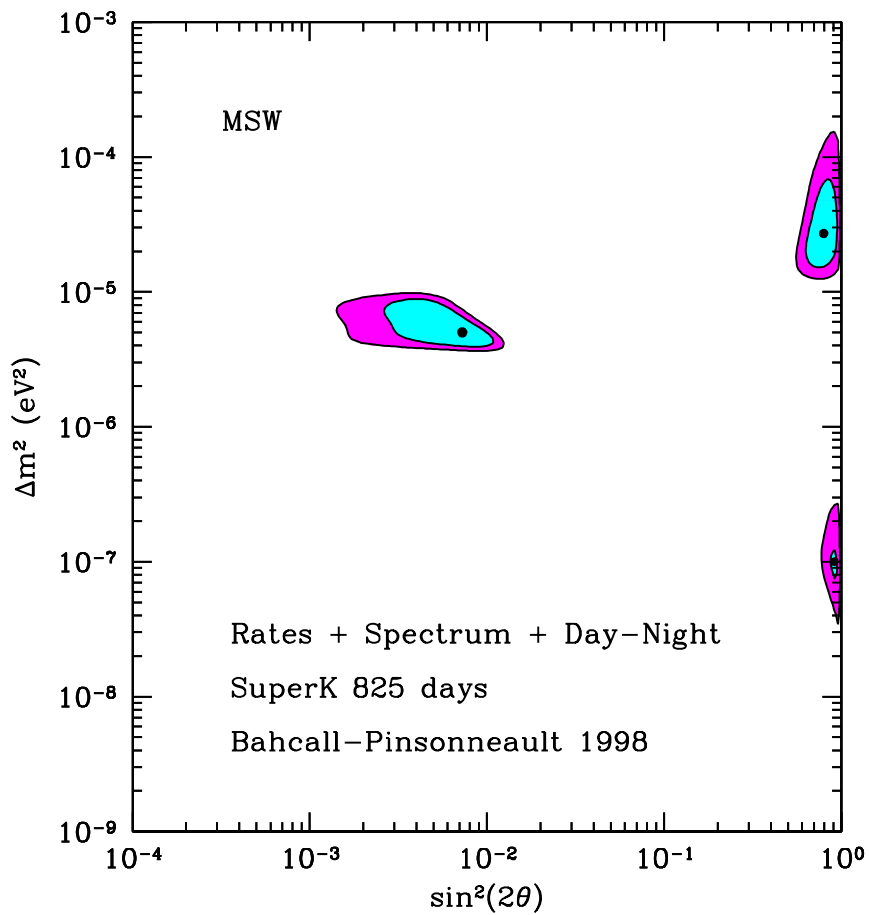}
\caption{\small{Allowed two-flavour MSW solutions to the solar neutrino problem. The contours correspond to 90\percent\ and 99\percent\ confidence limits. The input data comes from the total event rates in the Homestake, SAGE, GALLEX, and Super-Kamiokande experiments, as well as electron recoil energy spectrum and day-night effect data taken by Super-Kamiokande during 825 days of operation. The black dots within each region mark the position of the best-fit points in parameter space. (From~\cite{JNBhomepage}).}}
\label{MSW}
\end{center}
\end{figure}

The effect of oscillations on the observed rates in neutrino experiments can be represented on an \emph{MS diagram}, named after Mikheyev and Smirnov. An MS diagram defines a region of parameter space with orthogonal coordinates $\sin^{2}{2{\theta}}$ and ${{\Delta}m^2}$. Contours of constant event rate are plotted. The allowed solutions are enclosed within the contours. There are thought to be three allowed MSW solutions to the solar neutrino problem~\cite{JNB6}:
\begin{itemize}
\item The SMA (small mixing angle) solution\\
${{\Delta}m^2}$ = 5.0$\times$10$^{-6}$~eV$^2$\\
$\sin^{2}{2{\theta}}$ = 8.7$\times$10$^{-3}$
\item The LMA (large mixing angle) solution\\
${{\Delta}m^2}$ = 1.3$\times$10$^{-5}$~eV$^2$\\
$\sin^{2}{2{\theta}}$ = 0.63
\item The LOW (low probability, low mass) solution\\
${{\Delta}m^2}$ = 1.1$\times$10$^{-7}$~eV$^2$\\
$\sin^{2}{2{\theta}}$ = 0.83.
\end{itemize}
The SMA, LMA, and LOW solutions are shown on the MS diagram in figure~\ref{MSW}. LOW is an unlikely solution to the solar neutrino problem, but not excluded~\cite{Langacker}.
\newpage
\section{Evidence for Neutrino Oscillations}
Super-Kamiokande has found convincing evidence for neutrino oscillations, and therefore non-zero neutrino mass. If neutrinos do indeed have mass this will have profound consequences for astrophysics and cosmology because neutrinos might then make a significant contribution to the dark matter content of the universe~\cite{PPbook,JNBbook}. Neutrino oscillations also imply the existence of other non-trivial effects, such as non-conservation of lepton flavour number~\cite{RRichep}.

Super-Kamiokande detects atmospheric neutrinos coming from all directions. These neutrinos are created in the atmosphere all around the Earth. Super-Kamiokande measured the total flux of downward-going atmospheric neutrinos (coming from all directions above the horizontal at the detector) and the total flux of upward-going atmospheric neutrinos (coming from all directions below the horizontal at the detector). The up-down event ratio, $U/D$, measured at Super-Kamiokande for atmospheric muon neutrinos with energies greater than 1.33~GeV is 0.54; below unity by more than 6$\sigma$~\cite{nunews,SKichep}. In the absence of neutrino oscillations, one would expect the upward and downward fluxes to be equal.

\begin{figure}[!htbp]
\begin{center}
\epsfig{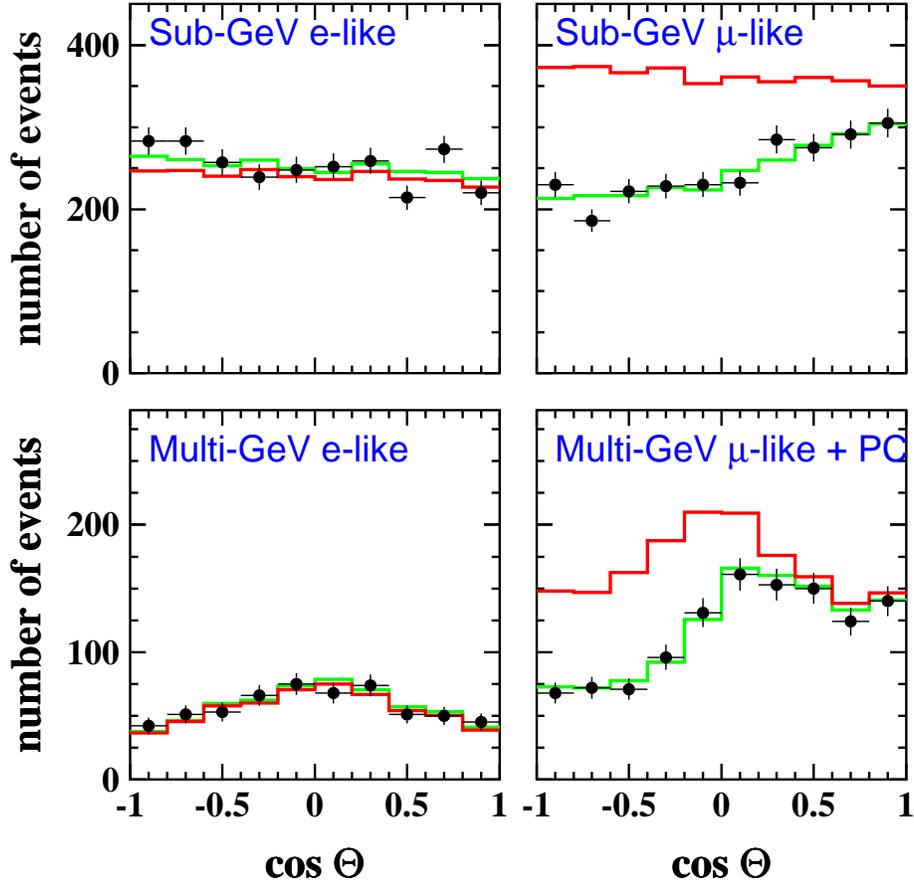}
\caption{\small{Zenith angle distributions for sub-GeV and multi-GeV electron-type and muon-type events at Super-Kamiokande. The red lines show the (no-oscillations) Monte Carlo predictions; the green lines show the predictions for \numu\goesto\nutau\ oscillations with ${\Delta}m^2$=3.2$\times$10$^{-3}$~eV$^2$ and $\sin^2{2\theta}$=1.0. (From~\cite{nunews}).}} % <-- References to colours here.
\label{zendep}
\end{center}
\end{figure}

Figure~\ref{zendep} shows the zenith angle distributions of electron-type (showering) and muon-type (non-showering) events observed by Super-Kamiokande. These events are subdivided into sub-GeV (energy $<$ 1.33~GeV) and multi-GeV (energy $>$ 1.33~GeV) events~\cite{nunews}. The zenith angle is measured from directly above the detector. The measured zenith angle distributions for electron-type events agree well with those predicted by Monte Carlo simulations, but the same is not true for muon-type events. The data exhibits a zenith angle dependent muon neutrino deficit which is inconsistent with predictions of atmospheric neutrino flux from (no-oscillations) Monte Carlo simulations~\cite{SKosc}. The data is consistent with a model in which upward-going muon neutrinos, which have travelled from the far side of the Earth, are depleted by oscillations.

\begin{figure}[!htbp]
\begin{center}
\epsfig{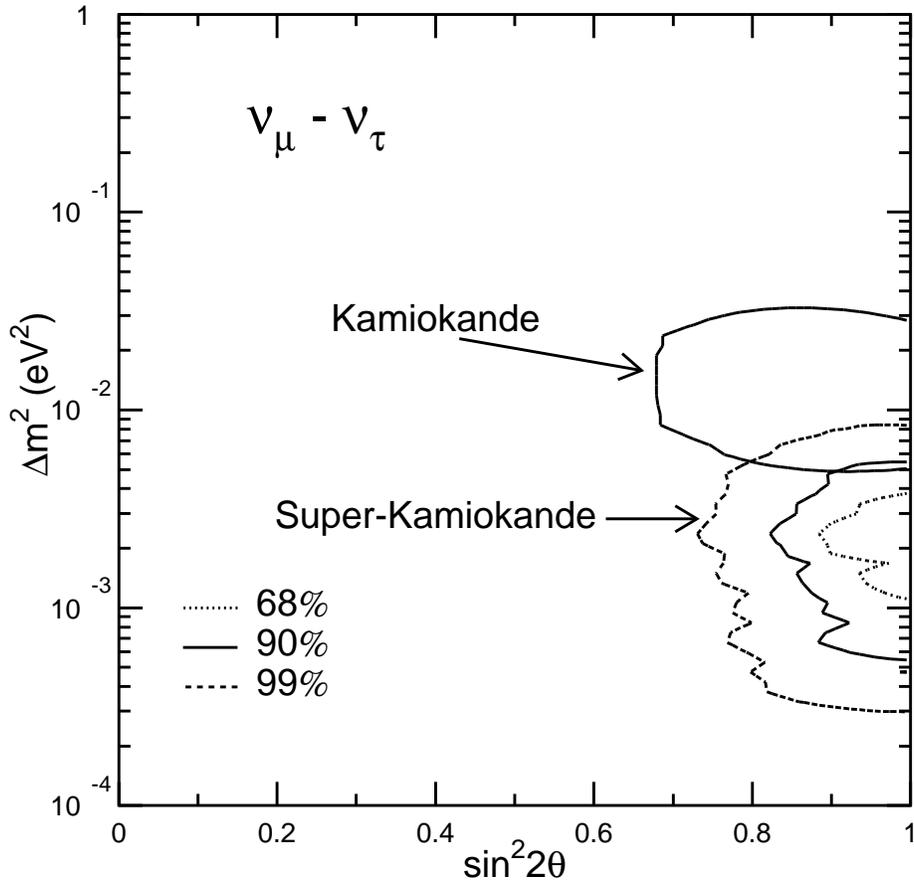}
\caption{\small{MS diagram for two-flavour \numu\goesto\nutau\ oscillations, based on data from Super-Kamiokande. 68\percent, 90\percent\ and 99\percent\ confidence intervals are shown for $\sin^2{2\theta}$ and ${\Delta}m^2$. Overlaid is the 90\percent\ confidence interval obtained by the Kamiokande~II experiment. (From~\cite{SKosc}).}}
\label{MSSK}
\end{center}
\end{figure}

The data is in good agreement with two-flavour \numu\goesto\nutau\ oscillations with \mbox{$\sin^2{2\theta}>$ 0.82} and \mbox{5$\times$10$^{-4} < {\Delta}m^{2} <$ 6$\times$10$^{-3}$~eV$^2$} at 90\percent\ confidence level~\cite{SKosc}. Figure~\ref{MSSK} shows the Super-Kamiokande results on an MS diagram. The allowed region of parameter space obtained from the Kamiokande~II experiment is overlaid.

The Super-Kamiokande Collaboration have excluded two-flavour \numu\goesto\nuster\ oscillations at the 99\percent\ confidence level as a solution to the atmospheric neutrino anomaly~\cite{SKosc,nunews}. \mbox{Two-flavour} \numu\goesto\nue\ oscillations are also excluded as a solution to the atmospheric neutrino anomaly, because if this channel was dominant it would lead to a distortion of the zenith angle distributions of electron-type events, contrary to observations (see figure~\ref{zendep})~\cite{nunews,SKichep}. Two-flavour \numu\goesto\nue\ oscillations are also excluded by CHOOZ, a long-baseline reactor experiment, as the main channel of atmospheric neutrino oscillations~\cite{SKichep}.

\section{Outlook}
Neutrino physics is now a burgeoning and very active field, and a large number of neutrino experiments are either scheduled to start in the very near future or are in advanced stages of planning. The SNO experiment, described earlier, will provide valuable insights into the solar neutrino problem by virtue of its ability to measure the total neutrino flux, of all flavours, from the Sun~\cite{SNO,TNG}. BOREXINO is a liquid scintillation experiment that will be able to measure the solar \Ber\ neutrino flux directly. This is an important measurement, because the four first-generation solar neutrino detectors (Homestake, SAGE, GALLEX, and Kamiokande~II) suggest that the \Ber\ neutrino flux is much less than that predicted by the standard solar model~\cite{TNG}. These experiments will be complemented by a number of reactor and long-baseline experiments. The first long-baseline experiment, K2K (KEK to Kamioka), has already started taking data, and MINOS (another long-baseline experiment) is under construction~\cite{Akhmedov}. There are also proposals for muon storage rings that would produce beams of muon neutrinos for use in future long-baseline experiments~\cite{Akhmedov}. 

These experiments, and theoretical work that is conducted in parallel, will bring us important knowledge about neutrinos that should allow us to answer many questions, solve the solar neutrino problem, and perhaps provide clues about physics beyond the Standard Model.

%%%%%%%%%%%%%%%%%%%%%%%%%%%%%%%%%%%%%%%%%%%%%%%%%%%%%%%%%%%%%%%%%%%%%%%%%%%%%%%

\end{document}